\vsize=23.5truecm \hsize=16truecm
\baselineskip=0.5truecm \parindent=0truecm
\parskip=0.2cm \hfuzz=1truecm

\font\scap=cmcsc10

\newcount\eqnumber
\eqnumber=1
\def\neweq{{\rm{(\the\eqnumber)}}\global\advance\eqnumber by 1}
\def\eqdef#1{\eqno\xdef#1{\the\eqnumber}\neweq}
\def\newaeq{{\rm{(\the\eqnumber a)}}\global\advance\eqnumber by 1}
\def\eqdaf#1{\eqno\xdef#1{\the\eqnumber}\newaeq}
\def\eqdisp#1{\xdef#1{\the\eqnumber}\neweq}
\def\eqdasp#1{\xdef#1{\the\eqnumber}\newaeq}

\newcount\refnumber
\refnumber=1
\def\newref{{\the\refnumber}\global\advance\refnumber by 1}
\def\refdef#1{{\xdef#1{\the\refnumber}}\newref}

\newcount\fignumber
\fignumber=1
\def\newfig{{\the\fignumber}\global\advance\fignumber by 1}
\def\figdef#1{{\xdef#1{\the\fignumber}}\newfig}

\def\smallskip{\vskip 3pt}
\def\medskip{\vskip 6pt}
\def\bigskip{\vskip 12pt}

\input graphicx.tex

\centerline{\bf The HANDY model with non-renewable resources and societal effects}
\bigskip
\medskip{\scap Mathilde Badoual}, {\scap Basil Grammaticos} and {\scap St\'ephane Plaszczynski}
\quad{\sl Universit\'e Paris-Saclay and Universit\'e de Paris-Cit\'e, CNRS/IN2P3, IJCLab, 91405 Orsay, France}
\medskip{\scap Ralph Willox} \quad
{\sl Graduate School of Mathematical Sciences, the University of Tokyo, 3-8-1 Komaba, Meguro-ku, 153-8914 Tokyo, Japan; willox@ms.u-tokyo.ac.jp}

\bigskip
{\sl Abstract}\par
We present an extension of the single-population Human and Nature Dynamics (HANDY) model that explicitly incorporates non-renewable resources and societal effects. The model is formulated as a system of four differential equations for which we provide a faithful discrete integrator. In the model, resource extraction and conversion is described by realistic Holling-type functional responses that capture saturation effects for both renewable and non-renewable resources, while efficiency losses are introduced to describe inherent waste. We derive analytic conditions on the model parameters, needed to ensure stable (and sustainable) equilibria, despite population growth and depletion of resources. The different dynamical scenarios that appear in this analysis and the parametric conditions that lead to them, are illustrated by extensive numerical simulations.
The societal effects we investigate include adjustable extraction and consumption rates of the resources, alongside a population-driven degradation of Nature and of its regenerative capacity. One immediate conclusion is that permanent environmental damage inevitably precipitates population collapse. In contrast, we also show that the possibility of recoverable capacities, even delayed, may lead to long-term survival and that timely late-stage interventions often succeed in averting extinction entirely.

\bigskip
MSC2010 numbers: 37M99, 39A30, 91C99\par

Keywords: population dynamics, collapse, resources and reserves, societal effects\par

Acknowledgements: RW gratefully acknowledges financial support from Arithmer Inc. through a collaborative research grant.
\bigskip

1. {\scap Introduction}
\medskip
Living organisms are mortal; that is the only certainty of biological existence. The lifespan of individual organisms has been studied extensively, especially in humans, where such knowledge gave rise to a  uniquely human institution that tames uncertainty by monetising risk: insurance. However, when it comes to entire species, the duration of their existence is far harder to measure, and in any case bears little to no relation to the lifespan of their individual members. Paleontology examines the history of species that inhabited Earth before our time and has identified (at least) five major events of mass extinction [\refdef\raupsep,\refdef\bigfive]. Some of these wiped out not only species but entire genera, families, and even orders. Most appear to have been triggered by climatic upheavals, though one, the Cretaceous extinction, is commonly attributed to an exogenous cause: the impact of a massive asteroid that provoked a global cataclysm and a sudden cooling of the atmosphere [\refdef\bolide]. Today, environmentally conscious observers speak of a “sixth mass extinction” [\bigfive], set in motion by human activity. This issue, however, is too contentious to be treated here and, instead, we shall confine ourselves to a deliberately anthropocentric question: the survival prospects of our current way of life and of the current structure of human society, and, as a related matter, the fate of the human species.

Human institutions and societal structures are no less mortal than the individuals who create them, though their lifetimes can be much longer. A glance at the great empires of the past reveals that, sooner or later all will have vanished,  with their lifespan only occasionally exceeding a millennium (as is thought to be the case for the Kushite Empire in Eastern Africa and, most notably, the Pandya Dynasty in South India). In any case, the age of empires is largely behind us. Society has evolved towards a global, interconnected civilisation, an all-encompassing human system, and the question we therefore wish to address extends to the survival, or potential collapse, of civilisation as we know it today.

For the better part of two centuries now, already since Malthus' pioneering work on population dynamics  [\refdef\malthus], the main point of concern in this respect has been the sustainability of exponential population growth, through an intense use of non-renewable resources. While demographic projections point towards a ``peak population'', to be attained later in this century [\refdef\lancet], the rate of depletion of resources should caution against too much carefree optimism. Faced with the possibility of an anthropogenic extinction of the human species, several authors have sought an answer through the framework of mathematical modelling. Forrester was probably the first to address the question of the survival of human society in a systematic way in an approach dubbed ``World Dynamics'' [\refdef\forrester]. He identified the two major dangers facing modern society: over-population and disparity in the standards of living (with pollution playing also a non-negligible role). He proposed a mathematical model of vast scope that he---modestly---called ``World''. This model was an evolution of Forrester's previous models on Urban Dynamics [\refdef\preforrester] and it included effects of population, capital investment, geographical space, natural resources, pollution and food production. The ``World'' model rose to the limelight through the collaboration of Forrester with the team of Meadows and their association to the Club of Rome, who had already planned a work on the ``Predicament of Mankind'' [\refdef\predicament], a kind of modern-era heir to Malthus' ``Essay on the principle of population''. The ``Limits to Growth'' [\refdef\limits] monograph of Meadows and collaborators, based on Forrester's World-3 model, brought home the message that the world's economy tends to stop its growth and collapse as the result of a combination of reduced resource availability, overpopulation, and pollution. Understandably, this sparked strong reactions. Eschewing all political considerations, siding with the Forrester-Meadows detractors we can point out that the model involves hundreds of variables, for which one must specify initial conditions, and thousands of parameters which have to be validated. Knowing the former with sufficient precision and fixing the latter in a realistic way is a nigh-impossible task. This simple criticism suffices to cast considerable doubt on the robustness of any long-term predictions made by means of that type of model. On the other hand, supporters of the methodology of ``Limits to Growth'' [\refdef\supporters] point out that most of the runs of the model, under varying assumptions, invariably lead to a medium-to-long term collapse [\refdef\jackson]. 

At the antipode of the immense complexity of the World Dynamics model, one finds the work of Brander and Taylor [\refdef\brandertaylor] who set out to model the economy of Rapa Nui (Easter Island) with a very simple model with just two variables, population and resources, and two parameters which control the population growth using the existing resources. The model itself is based on the hypothesis of a Malthusian population growth, relying on an open-access renewable resource and a simple Ricardian production structure (which is why Brander and Taylor refer to their model as a `Ricardo-Malthus'-type model). The purpose of the model was to offer an explanation for a purported massive population decrease on Rapa Nui, supposed to have occurred in the 1600s after (and to have been provoked by) a very prosperous but excessively exploitative period of economic activity. Although this characterization of Rapa Nui's economy and the conjectured population decline, long presented as established theory and popularized through bestsellers such as [\refdef\diamond], has been proven wrong in recent years [\refdef\rapanuipop,\refdef\rapanuigen], the model itself does have merit in that, with minimal parametric freedom, it succeeds in describing both a collapse as well as a steady state type of outcome. A model identical to that of Brander and Taylor, albeit proposed in a totally different context, was analysed in a previous work of some of the present authors [\refdef\cryptic]. 

Roughly ten years ago a paper by Motesharrei and collaborators was published addressing the issues of sustainability and collapse, while laying emphasis on societal disparities [\refdef\originalhandy]. They proposed a very simple model, which they dubbed ``HANDY'', again involving just two populations, ``elites'' and ``commoners'', in which the elites practically monopolise all the wealth produced by the commoners, leaving them only with a very small part that barely exceeds the subsistence minimum. Motesharrei et al. concluded that two different modes of collapse existed, a predictable one involving the depletion of natural resources and a different one due to the economic stratification, where the rapacity of the elites lead to starvation and disappearance of the commoners, resulting into total collapse. As could be expected, the paper, due to its inherent political overtones, lead to strong criticisms and it took some years before anybody would venture into the study of what was undisputedly an interesting model. Two of the present authors, in collaboration with J. Satsuma,  decided to do just this while stripping the model of the societal disparity component. In [\refdef\notrehandy] we proposed a single-population model, based on that of [\originalhandy], studied its dynamical properties and performed a few exploratory simulations with the help of a discrete version that we proposed and which we used as a faithful integrator. The results of these explorations were that, depending on the parameters, the society could be led to a global collapse or to a steady state but it turned out that in some special cases a limit cycle with alternating periods of prosperity and scarcity could exist as well. 

However, the attraction of the two-population model, resonating with the current structure of our society, finished by drawing several authors to the study of the original HANDY model (all the more so since by then the reactions to the initial publication had largely subsided). In a paper with the intriguing French-style subtitle ``A differential equations model without differential equations'' [\refdef\akhavanyorke],  Akhavan and Yorke revisited the HANDY model and studied the conditions needed for a population collapse to occur. One interesting extension of the original model is what they call ``downward mobility''. By allowing part of the elite population to revert to commoner status, they have shown that it is sometimes possible to avert a collapse and to sustain a stable society. Shilor and Kadhim [\refdef\shilorkadhim] have also studied a HANDY-like model with social mobility and have confirmed the predictions of Akhavan and Yorke through extensive numerical simulations. Tonnelier has studied the asymptotic states of the standard HANDY model using bifurcation analysis and has shown that in the same model a sustainable equilibrium  can coexist with cycles of prosperity and scarcity [\refdef\tonnelier]. A slightly different path was explored by Patry and collaborators [\refdef\patry] who consider the effect of exogenous perturbations, like natural disasters or wars, on the stability of the HANDY asymptotic states. Their conclusion is that the latter are rather robust since small perturbations do not suffice to radically modify them, while, they argue, large perturbations are rare in the real world (although this clearly depends on the definition of ``large'').

In this paper we introduce and study an extension of our single-population HANDY model in which we not only consider the effect of non-renewable resources  (which was not treated in a fully satisfactory way in [\notrehandy]) but in which we also consider the effect that the population has on Nature and its possible impact on sustainability. Some further extensions of the ``standard'' form of the HANDY model will also be introduced but these are best explained on the model itself, which will be done in Sec. 2 where we define the model and analyse its dynamical properties. In Sec. 3 we provide a discrete integrator for the model, which we use in Sec. 4 for a preliminary numerical exploration of the different asymptotic regimes that are attainable in the model. In Sec. 5 we analyse the potential influence of society on the evolution of the population and the resource reserves, again using numerical simulations. Finally, Sec. 6 is devoted to the conclusions.

\bigskip
2. {\scap Extending the single-population HANDY model}
\medskip
The initial single-population HANDY model we considered in [\notrehandy] involved three interdependent variables: the `population' $x$, the `(renewable) resources' $y$ and the `reserves' $z$. It is clear what we understand by `population' but we insist on the fact that it consists of a single, unique, societal class that participates equally in production and consumption. The renewable resources can be replenished by Nature up to a maximal capacity, which, as we shall see, may depend on the population. The means of subsistence of the latter are drawn from Nature through work, any surplus being added to the reserves. In [\notrehandy] we made the assumption that pre-existing non-renewable resources could be thought of as part of the reserves by introducing an appropriate initial condition for $z$. It turns out that this is not an optimal choice (as we shall explain shortly) and that for a more realistic approach we must introduce an additional variable, $w$, specifically for the non-renewable resources.  An extra variable, $r$, which plays a crucial role in the dynamics, is defined in terms of the fundamental ones: it represents the reserves per capita, i.e. $r=z/x$. 

Having introduced the variables we can now write the equations for our extended HANDY model:
$${dx \over dt}=A(r)x\eqdaf\zdyo$$
$${dy \over dt}=F(x,y)-D(x,y)\eqno(\zdyo \rm b)$$
$${dz \over dt}=D(x,y)+E(x,w)-B(r)x\eqno(\zdyo \rm c)$$
$${dw \over dt}=-E(x,w).\eqno(\zdyo \rm d)$$
The population equation (\zdyo a) is just a Malthusian one where the growth is governed by the availability of reserves (per capita). In [\notrehandy] we introduced the convenient parametrisation $A(r)=\alpha -\beta\exp(-\lambda r)$, where $\beta>\alpha>0$ and $\lambda>0$, meaning that when the per capita reserves $r$ become scarce, the population starts dying out. 

The `Nature' (renewable resources) equation (\zdyo b) comprises two terms. The first governs the evolution of Nature while the second corresponds to the extraction of resources. The initial HANDY assumption for the first term is a simple logistic-type expression of the form $F(x,y)=y(C(x)-\kappa y)$ for some constant $\kappa>0$. The extension we introduce at this point is to allow the maximal capacity $C(x)$ for the renewable resources to depend on the population, i.e. $C(x)$ is a (positive) decreasing function of $x$. This incorporates not only  the effects of human habitats encroaching on arable land, but also that of pollution, degrading the environment and hindering its regeneration. When we ignore such effects $C(x)$ is just a (positive) constant. We shall return to the discussion of the capacity term later in this paper.
For the second term, the extraction term  $D(x,y)$, the form used in the original HANDY paper (as well as in [\cryptic]) was a simple (uniform) mixing  term  $xy$. Given the fact that the basic inspiration for this depletive term is coming from a predator-prey model, we decided to revisit it in the spirit of the predation theories of Holling. In [\refdef\holling] Holling argues that the rate of consumption of prey by the predator (what he calls the functional response) cannot increase indefinitely with the prey population. The rationale being that the predator can only explore a finite domain in search of its prey and does not have access to the total prey population. He formalised these arguments by introducing what he called the ``disk equation'' whereupon the functional response is of the form $PN/(1+\sigma N)$ where $P$ and $N$ are the populations of predator and prey respectively. Transposing the philosophy of Holling to the case of the extraction of resources we propose the term $D(x,y)=hxy/(x+y)$ instead of the simple product $xy$. This seems quite reasonable as for a very large population ($x\to\infty$) we have $D(x,y)\approx hy$ and the extraction is essentially proportional to the totally available quantity of resources, whereas in case of abundant resources ($y\to\infty$) we have $D(x,y)\approx hx$, i.e. the extraction is  proportional to the number of persons tackling the task. Note that we could have added a coefficient in front of any term of the denominator but it is clear that it can be absorbed through a proper scaling. Also, since there is a possibility that both $x$ and $y$ tend to zero at approximately the same speed, and in view of the numerical simulations we will perform, we opted to regularise the $xy/(x+y)$ term by adding a vanishingly small (positive) constant $\varepsilon$ to the denominator. We must emphasize however that this is merely a numerical trick and that the precise value of $\varepsilon$ does not influence any of the stability results we will derive hereafter, as long as $\varepsilon$ is chosen sufficiently small.

We apply the same Holling-inspired prescription (and subsequent regularisation) to the extraction of non-renewable resources, i.e. $E(x,w)=gxw/(x+w)$. The term $-B(r)x$ in (1c) describes the use of reserves, which can exceed the necessities for mere subsistence and can increase in a situation of prosperity. In [\notrehandy] we introduced the parametrisation $B(r)=\gamma -\delta\exp(-\mu r)$ where $\gamma>\delta>0$ and $\mu>0$, meaning that a subsistence minimum is a baseline necessity.

We can now write the equations of our version of the extended HANDY model as
$$x'=A(r)x\eqdaf\zdyo$$
$$y'=y(C(x)-y)-{hxy\over x+y+\varepsilon}\eqno(\zdyo \rm b)$$
$$z'={hxy\over x+y+\varepsilon}+{gxw\over x+w+\varepsilon}-B(r)x\eqno(\zdyo \rm c)$$
$$w'=-{{\tilde g}xw\over x+w+\varepsilon},\eqno(\zdyo \rm d)$$
where all parameters are taken to be non-negative.
Note that when $C(x)=C$ (constant), rescaling the time variable (which is the last remaining scaling freedom of the above system) allows us to put $C=1$ in all generality.
The rates of use and extraction, $g$ and $\tilde g$, of the non-renewable resources are not necessarily the same but obey $\tilde g\geq g$. We allow, in fact, for some inefficiency in the use of these resources, where not all that is extracted becomes a `reserve'. (We could have choosen the scaling so as to introduce different rates of use and extraction, $h$ and $\tilde h$, for the renewable---instead of the non-renewable---resources, but decided against this in order to keep the model without non-renewable resources as close as possible to the initial HANDY model).

Since the per capita reserves variable $r$ is an essential one, entering explicitly through $A(r)$ and $B(r)$, it is convenient to replace equation (2c) by an equation for this (intensive) variable. We find 
$$r'={hy\over x+y+\varepsilon}+{gw\over x+w+\varepsilon}-B(r)-rA(r).\eqno(\zdyo \rm e)$$
Given the form of equation (\zdyo c), or of its intensive variant (\zdyo e), it cannot be excluded a priori that the evolution described by these equations might lead to negative values for $z$ (or $r$). For our original modification of the HANDY model given in [\notrehandy] we took great care to identify parameter regimes that would preserve positivity for all variables and, likewise, adopted a special positivity preserving numerical scheme. Here we choose a different approach.  We allow for all possible (non-negative) parameter values in the model, but fix the values of $z(t)$ (and $r(t)$) at the value  $0$ if they would happen to pass through zero at some time $t_\star$. We then pursue the evolution of $x(t), y(t)$ and $w(t)$ with $z(t)=r(t)=0$ in (\zdyo a), (\zdyo b) and (\zdyo d) until equations (\zdyo c) or (\zdyo e) would again lead to a positive value for $z(t)$ or $r(t)$ at some time $t'>t_\star$ at which point we resume the normal evolution of system (\zdyo). Since (\zdyo c) and (\zdyo e) describe the production of goods (`reserves') from renewable and non-renewable resources, from a modelling perspective it is completely satisfactory to have production cease during a time interval such as $[t_\star,t']$ during which production is physically impossible. Moreover, it should be clear that this alternative evolution rule is especially simple to implement on a discrete version of the model, which we shall discuss in Sections 3 and 4.

Having established the equations of our model we can now proceed to the study of the fixed points and their stability. Here, for the time being, we neglect the effects of population on the capacity of Nature but choose to keep the symbol $C$ for future reference. Equation (\zdyo a) readily  leads to two possibilities, either $x_*=0$ or $x_*\ne0$ in which case the term $A(r)$ must vanish, leading to $r_*=r_0={1\over\lambda}\log{\beta\over\alpha}>0$, i.e. the value of $r$ for which we have $A(r_0)=0$. 

Pursuing first the branch $x_*=0$, which corresponds to a collapse of the population, we have from (\zdyo b) either $y_*=0$ or $y_*=C$. Following the latter value it is clear that (\zdyo d) does not constrain the value of the fixed point of $w$, which can be zero or positive: $w_*\geq0$. From (\zdyo e) we then obtain a condition, at leading order in $\varepsilon$, which fixes the value of $r$ at the fixed point: $r=r_*$ such that either $h+g=B(r_*)+r_*A(r_*)$ if $w_*\neq0$ or $h=B(r_*)+r_*A(r_*)$ if $w_*=0$, up to vanishingly small contributions in $\varepsilon$. Similarly,  when $x_*=y_*=0$ we find that the condition that fixes the value of $r_*$ is now $g=B(r_*)+r_*A(r_*)$ if $w_*\neq0$ or $B(r_*)+r_*A(r_*)=0$ when $w_*=0$, again up to vanishingly small contributions in $\varepsilon$.

The case $x_*\ne0$ (which would signal the absence of collapse) is more interesting. From (\zdyo d) we see immediately that $w_*=0$. Equation (\zdyo b) leads to two possibilities, either $y_*=0$ or $C=y_*+hx_*/(x_*+y_*)$. The former case can be discarded since it would lead, in (\zdyo e) to $B(r_0)=0$, which is impossible since $B(r)$ never vanishes. Thus we are left with the latter choice of a non-zero $y_*$ in which case we find from (\zdyo e) the condition $hy_*/(x_*+y_*+\varepsilon)=B(r_0)$. The values of $x_*,y_*$ can be easily obtained in terms of $B_*\equiv B(r_0)$: $x_*=(B_*+C-h)(h-B_*)/B_*$ and $y_*=B_*+C-h$, at leading order in $\varepsilon$. These coordinates are positive (in order to have physical meaning) only when
$$C+B_*>h>B_*.\eqdef\ztri$$

The stability of the `collapse' fixed point $x_*=0$ is straightforward to study. For $y_*=C$, the roots of the characteristic equation we obtain when linearising around this fixed point are obtained almost by inspection. We find the 4 roots $A(r_*), -C, -B'(r_*)-A(r_*)-r_*A'(r_*)$ and $0$, which do not depend on $\varepsilon$ at all. For the `collapse' fixed point to be attractive we must have $A(r_*)<0$. (Clearly, unless $A(r)<0$ the population will keep growing). Hence $r_*$, for all practical purposes obtained from  $h+g=B(r_*)+r_*A(r_*)$ (or $h=B(r_*)+r_*A(r_*)$ when $w_*=0$), must first of all satisfy the condition: $0\leq r_*<r_0$. Moreover, this value should be such that ${d\over dr}\left(B(r)+r A(r)\right)|_{r=r_*}>0$. Thus we have the collapse condition $0>A(r_*)>-B'(r_*)-r_*A'(r_*)$, provided such an $r_*$ exists. This is definitely possible since  the function $B(r)+r A(r)$ necessarily has a minimum at some $r=r_1$ in the half-closed interval $[0,r_0[$. Therefore, if $h+g$ (or $h$, depending on the value of $w_*$) lies in the interval $[B(r_1)+r_1 A(r_1),\gamma-\delta e^{-\mu r_0}[$ then there exists a unique $r_*\in [r_1,r_0[$ that satisfies the above conditions.

The case $y_*=0$ leads to the same set of roots for the characteristic equation with one essential exception: the second root, associated with the stability of $y_*=0$, is $+C$ instead of $-C$.  This shows that the value $y=0$ cannot be reached asymptotically and that although the natural resources might become severely depleted they cannot vanish entirely and, sooner or later, will rebound and might even reach full capacity again if in the meanwhile the population has vanished and there is no longer any extraction. (We shall illustrate this behaviour through numerical simulations in the next section). 

Next we turn to the `steady state' fixed point $x_*\ne0$. In order to condense the notations we introduce $A_*'\equiv A'(r_*)$ and similarly for $B$. To investigate the stability of the `steady state' we derive the linearised version of (\zdyo) around the fixed point and compute the corresponding characteristic polynomial. At leading order in $\varepsilon$, the latter factorises into a linear and a cubic part. The linear part gives the root $-\tilde g$ which corresponds to the attraction to $w_*=0$. The cubic part is 
$$h\rho^3+(B_*^2+hC+hD_*-h^2)\rho^2+\big(A_*'B_*(h-B_*)+D_*(B_*^2+hC-h^2)\big)\rho+A_*'B_*(h-B_*)(C+B_*-h)=0,\eqdef\ztes$$
where $D_*=B_*'+r_*A_*'$. Next we invoke the Routh-Hurwitz stability criterion [\refdef\gantmacher] which gives a necessary and sufficient condition for the roots of the characteristic polynomial to have negative real parts. In the present case the criterion requires that all coefficients of the powers of $\rho$ in the cubic polynomial be positive and moreover that the product of the coefficients of $\rho$ and $\rho^2$ be larger than the product of the coefficients of $\rho^3$ and $\rho^0$. The latter translates into the condition
$$A_*'B_*(h-B_*)(B_*^2-hB_*+hD_*)-D_*(h^2-hC-B_*^2)(B_*^2+hC+hD_*-h^2)>0,\eqdef\zpen$$
while the positivity conditions on the coefficients yield two more constraints
$$hD_*>h^2-hC-B_*^2,\eqdef\zhex$$
and
$$A_*'>{D_*(h^2-hC-B_*^2)\over B_*(h-B_*)}.\eqdef\zhep$$
Condition (\zpen) needs further study. To begin with let us suppose that $B_*^2-hB_*+hD_*>0$. In that case it is possible to write condition (\zpen) as 
$$A_*'>{D_*(h^2-hC-B_*^2)(B_*^2+hC+hD_*-h^2)\over B_*(h-B_*)(B_*^2-hB_*+hD_*)},\eqdef\zoct$$
where, given (\ztri), the ratio of the positive terms $(B_*^2+hC+hD_*-h^2)/(B_*^2-hB_*+hD_*)$ is larger than 1. Thus (\zoct) is a stronger constraint for $A_*'$ than (\zhep) and thus supplants it. When $B_*^2-hB_*+hD_*<0$ the only possibility for (\zpen) to be satisfied is for $h^2-hC-B_*^2$ to be also negative. In that case we obtain for $A_*'$ the constraint
$$A_*'<{D_*\big|h^2-hC-B_*^2\bigr|(B_*^2+hC+hD_*-h^2)\over B_*(h-B_*)\bigl|B_*^2-hB_*+hD_*\bigl|},\eqdef\zenn$$
while (\zhep) would leave (the positive) $A_*'$ free.

From a practical point of view, if one wishes to reach a steady state, one must choose the values of $B_*$ and $C$ in accordance to (\ztri), i.e. $B_*<h$ and $C>h-B_*$. Then choose a value of $D_*$ obeying (\zhex) and calculate the quantity $B_*^2-hB_*+hD_*$.  If it is positive, use (\zoct) as a constraint for the value of $A_*'$. If it is negative then verify the sign of $h^2-C-hB_*^2$. If this sign is negative, then use (\zenn) as a constraint for the value of $A_*'$. If it is positive, then the `steady state' fixed point cannot be stable.
(Note that, even if one chooses $B(r)$ to be constant, in which case $B_*'=0$, and $D_*=r_*A_*'$, the constraints (\zhex) and (\zoct)/(\zenn) still make sense, the latter being a condition on $r_*$, while the former allows to obtain $A_*'$ from the knowledge of $D_*$ and $r_*$). 

The above analysis shows the possibility of having, besides the collapse scenario $x_*=0, y_*=C$, a stable steady state with $x_*\neq0$ as another possible outcome for the evolution described by our model (\zdyo). This begs the question as to what happens when the parameters are such that all fixed points become unstable. In Sec. 4 we will show, through numerical simulation, that in such a case the system is attracted to a limit cycle.

\bigskip
3. {\scap The discrete HANDY model}
\medskip
We now proceed to construct an explicit integrator by discretising the differential equations in the model. 
We start by discretising the independent (time) variable $t$ as $t_n=t_0+n\tau$, where $\tau>0$ is a time step, which is in principle taken to be small. We introduce the notation $x_n\equiv x(t_n)$ and so on for the remaining variables and discretise the time derivative as a forward difference: ${dx\over dt}\approx (x_{n+1}-x_n)/\tau$, and so on. Then, in order to avoid to the fullest extent the appearance of minus signs  in the resulting discrete evolution equations, we introduce appropriate staggerings for the variables as:
$${x_{n+1}-x_n\over\tau}=\alpha x_{n}-\beta x_{n+1} e^{-\lambda r_n}\eqdaf\zdek$$
$${y_{n+1}-y_n\over\tau}=y_n(C-y_{n+1})-{hx_ny_{n+1}\over x_n+y_{n}+\varepsilon},\eqno(\zdek b)$$
$${z_{n+1}-z_n\over\tau}={hx_ny_{n+1}\over x_n+y_{n}+\varepsilon}+{gx_nw_{n+1}\over x_n+w_{n}+\varepsilon}-\gamma x_n+\delta x_n e^{-\mu r_n},\eqno(\zdek c)$$
$${w_{n+1}-w_n\over\tau}=-{\tilde g x_nw_{n+1}\over x_n+w_{n}+\varepsilon},\eqno(\zdek d)$$
while for $r$ we have
$${r_{n+1}-r_n\over\tau}={hy_{n+1}\over x_n+y_{n}+\varepsilon}+{gw_{n+1}\over x_n+w_{n}+\varepsilon}-\gamma+\delta e^{-\mu r_n}-\alpha r_{n+1}+\beta r_n e^{-\lambda r_n}.\eqno(\zdek e)$$
Note that we have introduced a different staggering for each of the two components in $A(r)$, i.e. $\alpha x_n$ and $\beta\exp(-\lambda r_n)x_{n+1}$ in (\zdek a) as well as $r_{n+1}\alpha$ and $r_n\beta\exp(-\lambda r_n)$ in (\zdek e) so as to minimise the impact of negative signs. However when it comes to $B(r)=\gamma -\delta\exp(-\mu r)$ in (\zdek e) and (\zdek c), while the term proportional to $\delta$ enters with a positive sign, there is no staggering that enables us to avoid the negative sign coming form the $\gamma$ term. In [\notrehandy] we introduced a workaround, which guaranteed positivity at the price of non-reversibility for the evolution equations. We shall return to this point in Sec. 3 but here we decided to keep the negative term in (\zdek e) and work with the discrete system

$$x_{n+1} = {(1+\alpha\tau) x_n \over 1+\beta \tau e^{-\lambda r_n}},\eqdaf\zdekbis$$
$$y_{n+1} = {(1+c\tau) y_n \over 1+\tau y_n + {\tau h x_n \over \varepsilon + x_n + y_n}},\eqno(\zdekbis b)$$
$$w_{n+1} = {w_n \over 1 + {\tau \tilde g x_n \over \varepsilon + x_n + w_n}},\eqno(\zdekbis c)$$
$$r_{n+1} = {1\over 1+\alpha\tau} \left[ r_n -\tau\gamma + \tau\big( \delta e^{-\mu r_n} + \beta r_n e^{-\lambda r_n} + {g w_{n+1} \over \varepsilon + x_n + w_n} + + {h y_{n+1} \over \varepsilon + x_n + y_n} \big)\right].\eqno(\zdekbis d)$$

We claim that system (\zdekbis) is a faithful transcription of (\zdyo) to a discrete setting. By construction, the fixed points of (\zdekbis) are the same as those for the continuous system, so it suffices to verify that their stability properties are also the same.
Studying in full detail the stability of the discrete system (\zdekbis) would lead to intractable conditions in the case of the steady state fixed point, that would be completely useless in practice. Still it is instructive to do so in a simpler case, that of a collapse scenario with a fixed point at $x_*=0, y_*=C$, $w_*\geq0$ (but otherwise free), and $r_*$ such that $h+g=B(r_*)+rA(r_*)$ if $w_*\neq0$ or $h=B(r_*)+rA(r_*)$ if $w_*=0$, up to vanishingly small contributions in $\varepsilon$. In this case the roots of the characteristic polynomial obtained by linearizing (\zdekbis) around this fixed point, are
$${ 1 +\alpha \tau \over 1 + \beta \tau e^{-\lambda r_*}}\,,~~ {1\over 1+\tau C}\,,~~ { 1 + \tau\big(\beta(1-\lambda r_*) e^{-\lambda r_*} - \delta\mu e^{-\mu r_*}\big)\over 1+\alpha\tau}\,, ~~1,\eqdef\droots$$
i.e. again completely independent of $\varepsilon$.
For the stability of the fixed point of the discrete system all these roots of the characteristic polynomial should lie within the unit circle in the complex plane. 

First of all, the fourth root being equal to 1 signals a neutral stability of the fixed point value for $w$, similar to the differential system. The second root, $1/(1+\tau C)$, always lies between 0 and 1 as both $C$ and $\tau$ are positive and tells us that the $y$ values are always attracted to $y_*=C$, just as in the continuous case. The first root, $(1 +\alpha \tau)/(1 + \beta \tau e^{-\lambda r_*})$, is also always positive and requiring it to be less than 1 yields exactly the same constraint as in the continuous case:  $A(r_*)<0$. Finally, for the third root to be less than 1, the quantity $r_*$ must be such that $A(r_*) + r_* A'(r_*) + B'(r_*)>0$, again exactly as was the case for the continuous system (\zdyo). Note that, up to this point, all these conditions are independent of the discretisation step $\tau$. Requiring the third root to be greater than -1, however, does yield a $\tau$ dependent condition, which is nonetheless easily satisfied for small $\tau$. This is a general feature of discrete systems which are faithful discretisations of some differential system but do not preserve positivity for all possible time steps: for the same set of parameters, if the stepsize becomes too large, the stability properties of the fixed points in the discrete system might start to differ from those in the continuous system. 

\bigskip
4. {\scap Simulating the dynamical behaviour of HANDY}
\medskip
Having proposed an integrator for the model (\zdyo), we now turn to simulations to explore its possible dynamic behaviours. However, before proceeding, it is worthwhile summarising the various parameters entering our model as well as their physical meaning.

\bigskip
\font\large=cmr12 scaled \magstep2
\font\notso=cmr12 scaled \magstep1

\newdimen\thickw \thickw=1.5pt
\newdimen\thinw  \thinw=0.4pt
\newdimen\cellht \cellht=0.6cm

\newdimen\cellwdthick \cellwdthick=15cm
\advance\cellwdthick by -\thickw
\advance\cellwdthick by -\thickw

\newdimen\cellwdthin \cellwdthin=15cm
\advance\cellwdthin by -\thinw
\advance\cellwdthin by -\thinw

\vbox{
  \offinterlineskip
 \hbox{%
  \vrule width\thickw
  \vbox{
    \hrule height\thickw
    \vbox to1.5cm{\vss\hbox to\cellwdthick{\hss \hfil {\large The Model Parameters} \hfil \hss}\vss}
    \hrule height\thickw
  }%
  \vrule width\thickw
}%
   \hbox{%
    \vrule width\thinw
    \vbox{
      \vbox to 1cm{\vss\hbox to\cellwdthin{\hss \hfil {\notso Population} \hfil\hss}\vss}
      \hrule height\thinw
    }%
    \vrule width\thinw
  }%
  \hbox{%
    \vrule width\thinw
    \vbox{
     \vbox {\hbox to\cellwdthin{\hss \vbox{\hbox to 3cm {$A(r)=\alpha-\beta e^{-\lambda r}$\hfill}\vskip1.3cm } \includegraphics[width=4cm,height=3cm]{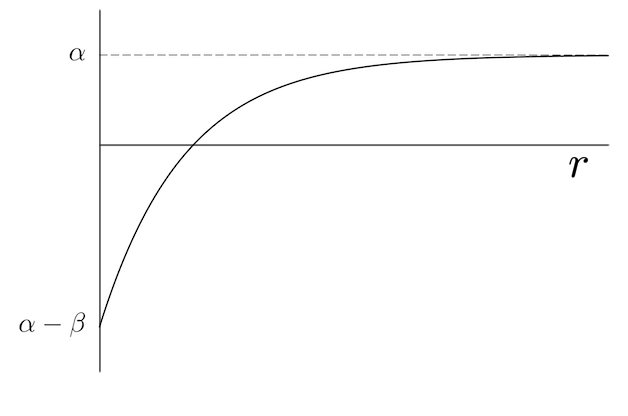}\qquad\vbox{
   \vskip 0.5cm
  \hbox to 3cm{$\alpha:$\ births\hfill}
  \vskip 0.5cm
  \hbox to 3cm{$\beta:$\ deaths\ $(\beta>\alpha)$\hfill}
\vskip 0.5cm
  \hbox to 3cm{$\lambda:$\ rate\hfill}
 \vskip 0.5cm
} \hss}}
      \hrule height\thinw
    }%
    \vrule width\thinw
  }%
  \hbox{%
    \vrule width\thinw
    \vbox{
      \vbox to 1cm{\vss\hbox to\cellwdthin{\hss {\notso Renewable Resources}\hss}\vss}
      \hrule height\thinw
    }%
    \vrule width\thinw
  }%
  \hbox{%
    \vrule width\thinw
    \vbox{
     \vbox {\hbox to\cellwdthin{\hss   \vbox{
   \vskip 0.5cm
  \hbox to 3cm{$F(x,y)=y(C(x)-y)$\hfill}
  \vskip 0.5cm
  \hbox to 3cm{$D(x,y)=h{xy\over x+y}$\hfill}
 \vskip 0.5cm
} \qquad\qquad
     \vbox{
   \vskip 0.5cm
  \hbox to 3cm{$C(x):$\ Nature capacity\hfill}
  \vskip 0.5cm
  \hbox to 3cm{$h:$\ extraction coefficient\hfill}
 \vskip 0.5cm
}\hss}}
      \hrule height\thinw
    }%
    \vrule width\thinw
  }%
  \hbox{%
    \vrule width\thinw
    \vbox{
      \vbox to 1cm{\vss\hbox to\cellwdthin{\hss {\notso Non-Renewable Resources}\hss}\vss}
      \hrule height\thinw
    }%
    \vrule width\thinw
  }%
  \hbox{%
    \vrule width\thinw
    \vbox{
     \vbox {\hbox to\cellwdthin{\hss \vbox {\hbox to 3cm {$E(x,w)={\tilde g}{xw\over x+w}$\hfill}\vskip 1.0cm } \qquad\qquad
     \vbox{
   \vskip 0.5cm
  \hbox to 3cm{${\tilde g}:$\ extraction coefficient\hfill}
  \vskip 0.5cm
  \hbox to 6cm{$g<{\tilde g}:$\ inefficient contribution to reserves\hfill}
 \vskip 0.5cm
}\hss}}
      \hrule height\thinw
    }%
    \vrule width\thinw
  }%
  \hbox{%
    \vrule width\thinw
    \vbox{
      \vbox to 1cm{\vss\hbox to\cellwdthin{\hss {\notso Consumption}\hss}\vss}
      \hrule height\thinw
    }%
    \vrule width\thinw
  }%
\hbox{%
    \vrule width\thinw
    \vbox{
     \vbox {\hbox to\cellwdthin{\hss \vbox{\hbox to 3cm {$B(r)=\gamma-\delta e^{-\mu r}$\hfill}\vskip1.3cm } \includegraphics[width=4cm,height=3cm]{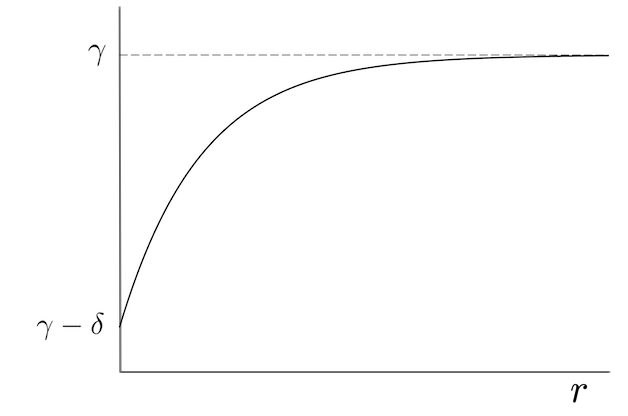}\qquad\vbox{
   \vskip 0.5cm
  \hbox to 3.8cm{$\gamma-\delta:$\ subsistence consumption\hfill}
  \vskip 0.5cm
  \hbox to 3.8cm{$\delta:$\ surplus consumption\ $(\gamma>\delta)$\hfill}
\vskip 0.5cm
  \hbox to 3cm{$\mu:$\ rate\hfill}
 \vskip 0.5cm
} \hss}}
      \hrule height\thinw
    }%
    \vrule width\thinw
  }%
}

In order to simulate the behaviour of HANDY we iterate the discrete system (\zdekbis) starting from appropriate initial conditions. In all cases presented here we took $x_0=10^{-3}$ as the initial condition for the population. We have checked that using a smaller value has the simple effect of delaying the onset of population growth, without affecting the subsequent dynamics. (This is an effect observed already in [\refdef\covid] concerning the onset of an epidemic). In the following we also always take the initial value for $y_0$ to be the capacity $C$. This is not a crucial choice, since the value of $y$ increases fast during the initial phase of the (rather slow) population growth, practically reaching capacity anyhow. In most simulations performed for this first part of the paper the capacity was taken to be constant, in which case it can be taken to be equal to 1 because of scaling freedom, as explained in Sec. 2.
The initial value for $r$ was chosen so as to have $A(r)=0$, thus preventing any initial bias towards initial decrease or increase of the population. (We have checked nonetheless that other choices of the initial value of $r$ do not influence the evolution of the system). The only quantity for which there is no a priori optimal choice is the initial value of the non-renewable resources $w$. We shall therefore fix an arbitrary initial value $w_0$ and investigate its effect on the evolution of the model. 

All simulations were carried out with a step $\tau=0.01$ (but, as in our initial HANDY publication [\notrehandy], we verified that changing $\tau$ by a factor of 10 either way does not alter the general behaviour) and $\varepsilon$, a few orders of magnitude larger than the numerical zero of the simulations. Since the simulations are extremely fast we decided to integrate also equation the equation for $z(t)$ using (\zdek c), and verified that the identity $z_n= r_n x_n$ holds at all time steps to a high precision. 
As mentioned in Sec.2, since there is an explicit negative term $-B(r_n)$ in (\zdek c), as well as in (\zdekbis d), there is no way to guarantee that the solution to these equations remains positive at all times. Our strategy in this case is the following: integrate for $z_n$ and $r_n$ till a negative value is reached. When this happens, put the value to 0 and pursue the integration, keeping the value at 0 till a positive value emerges. From then onwards proceed normally. The rationale behind this is that although the reserves are zero, it might be possible for the population to survive on what is being continuously produced from the renewable and non-renewable resources, at least for some time. Since the use of reserves led to their depletion, it is clear that what is being produced does not suffice in order to sustain the population and so the latter will decrease. Depending on the details of the model, this decrease may bring the population down to a level where reserves start being replenished again. 

In all the figures we shall present, the evolution of the population ($x$) is depicted in red, that of the renewable resources (i.e. ``Nature'', $y$) in green, the reserves ($z$) accumulated by the population in blue and the non-renewable resources ($w$) by a black dashed line. Since $w$ can only decrease over time, but as its initial value $w_0>0$ can be arbitrarily large compared to those of the other 3 variables, we shall always plot its rescaled value $w/w_0$ in order to increase the visibilty of the other evolutions in the figures. Whenever we consider non-contant Nature capacity we shall represent its evolution by a pink line.

Illustrating the influence of all model parameters through simulations would require dozens upon dozens of images. Therefore, despite the adage that “a picture is worth a thousand words,” we have chosen to present only the most essential figures, accompanied by comments on the parameters' effects, except when these effects markedly alter the dynamics, in which case we shall delve further into them.. 
As we saw in the previous section the system (\zdyo) possesses two regimes: population collapse and steady state, the transition between the two being governed by the set of inequalities from (\ztri) to (\zenn). We shall start our presentation of simulation results with the collapse scenario.

In the simulations that follow, the parameters in $A(r)$ are always the same: $\alpha=1, \beta=3$ and $\lambda=2$. Moreover, for the first  four figures (Figs. 1$\sim$4) we have chosen a $B(r)$ that describes a situation of constant consumption, i.e. in which $\delta=0: B(r) = \gamma$. Fig.  \figdef\one\ shows a typical collapse: 
\medskip
\centerline{\resizebox{10cm}{!}{\includegraphics{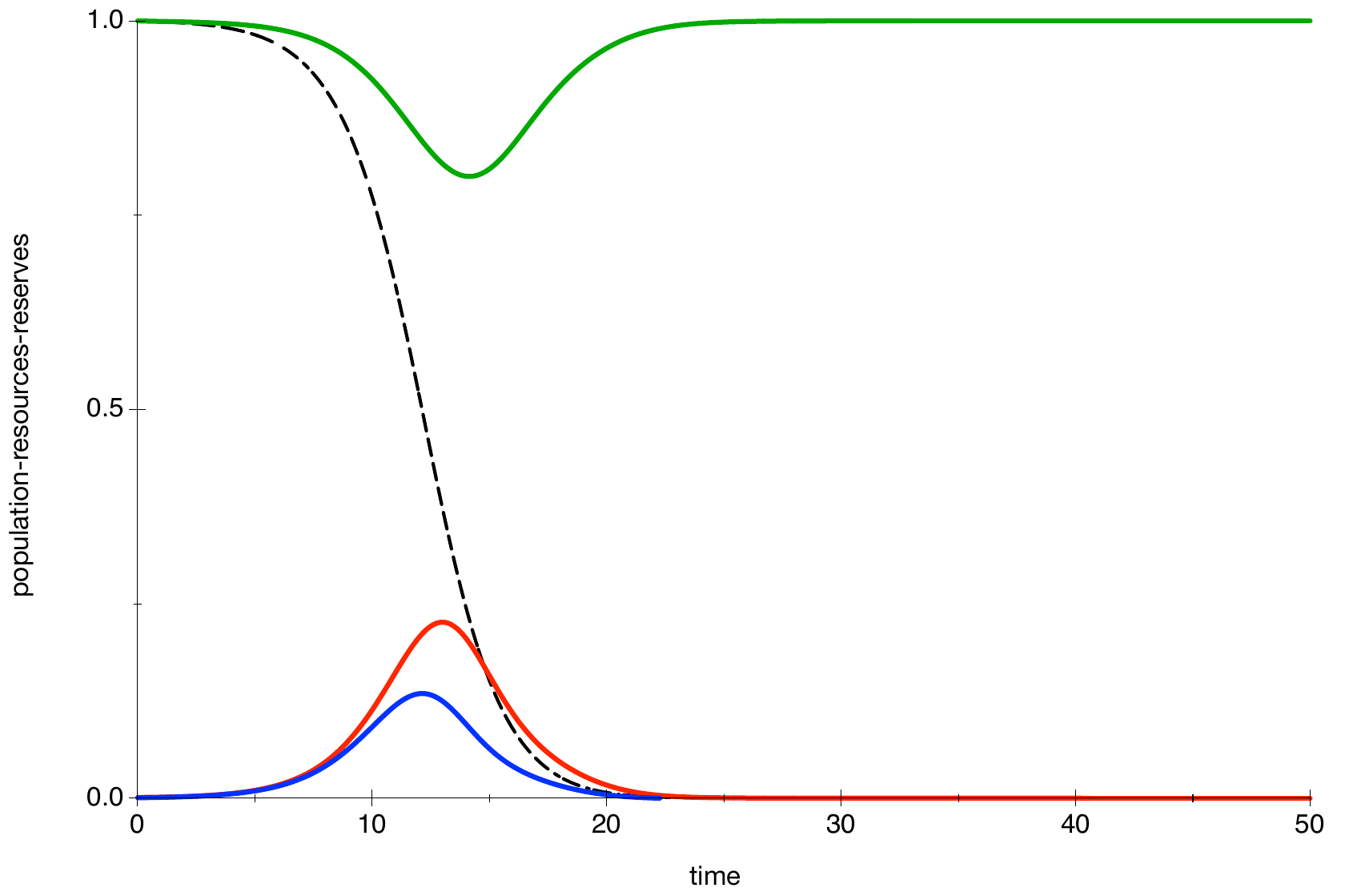}}}
\vskip-.25cm\qquad\centerline{\vbox{\hsize=14cm
\noindent Figure \one. A case of collapse. 
Simulation results for system (\zdyo) with parameter values $\alpha=1, \beta=3, \lambda=2, \gamma=1.5, \delta=0, c=1, g=\tilde g=1, h=1$, starting from initial conditions $x_0=10^{-3}, y_0=1, r_0= \log(\beta/\alpha)/\lambda$ and $w_0=1$. The plots show 5000 iterations for a time step $\tau=10^{-2}$.}}
\smallskip
Increasing $h$, but keeping always $\gamma>h$ (since the stability condition for the collapse fixed point with $w_*=0$ requires that $\gamma=h-A(r_*)>h$), only prolongs the pre-collapse period, with a temporary increase in the population and relative depletion of renewable resources. Increasing the value of $g$ leads also to a higher population but also to a more abrupt collapse, akin to the Seneca cliff [\refdef\bardi] we have identified in [\notrehandy], and to a total reserves depletion. 

The  effect of a larger initial value $w_0$ of $w$ is shown in Fig. \figdef\two. 
\medskip
\centerline{\resizebox{10cm}{!}{\includegraphics{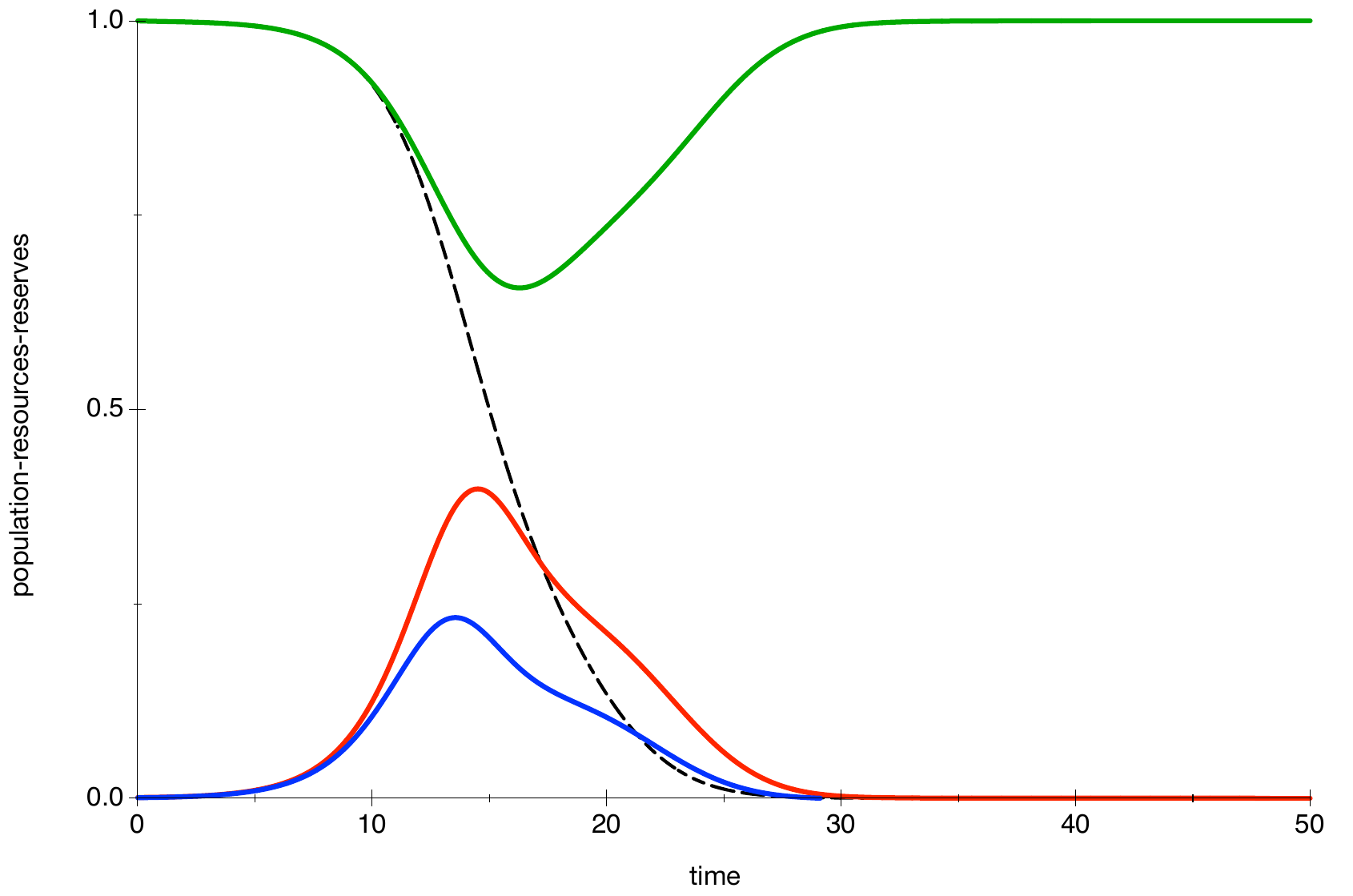}}}
\vskip-.25cm\qquad\centerline{\vbox{\hsize=14cm
\noindent Figure \two. Collapse with increased availability of non-neweable resources. 
 Simulation results for (\zdyo) with parameter values $\alpha=1, \beta=3, \lambda=2, \gamma=1.5, \delta=0, c=1, g=\tilde g=1, h=1$, starting from initial conditions $x_0=10^{-3}, y_0=1, r_0= \log(\beta/\alpha)/\lambda$ and $w_0=3$. The plots show 5000 iterations for a time step $\tau=10^{-2}$.}}
\smallskip
Again we are in a situation of collapse but the population reaches a higher maximum than when $w_0=1$ and the survival, i.e. pre-collapse, time is longer. In fact, much larger values of available non-renewable resources may lead to an illusion of a steady state, substantially prolonging the pre-collapse period, but as long as $\gamma>h$ collapse is inevitable. From here onwards we shall always work with $w_0=1$.

The effect of the birth and death terms ($\alpha$ and $\beta$) on a collapse situation like the one in Fig.  \one\ is as one can guess:  a small birth rate leads to a delayed increase of the population while a large one induces a rapid increase and a complete reserves depletion. 
The same effects are observed for a large and small death rate respectively. The value of the rate-parameter $\lambda$ also only plays a minor role, a small value leading to somewhat delayed population growth and the opposite in case of a larger value.

Decreasing progressively the value of $\gamma$ (but still always respecting $\gamma>h$) leads to a population exhibiting a long-tailed decay before eventually vanishing. Such a situation is shown in Fig. \figdef\three. We are still in a collapse situation but one which, for times corresponding to the long tail, might be perceived as a steady-state one.

The effect of inefficient extraction of non-renewable resources  (i.e. $g<\tilde g$)  is also not a determining one. Assuming that up to 30\% of what is extracted is lost, i.e. does not contribute to the reserves, we find that the population reaches a slightly lower maximum and that the survival time, till collapse, is shortened. 
\medskip
\centerline{\resizebox{10cm}{!}{\includegraphics{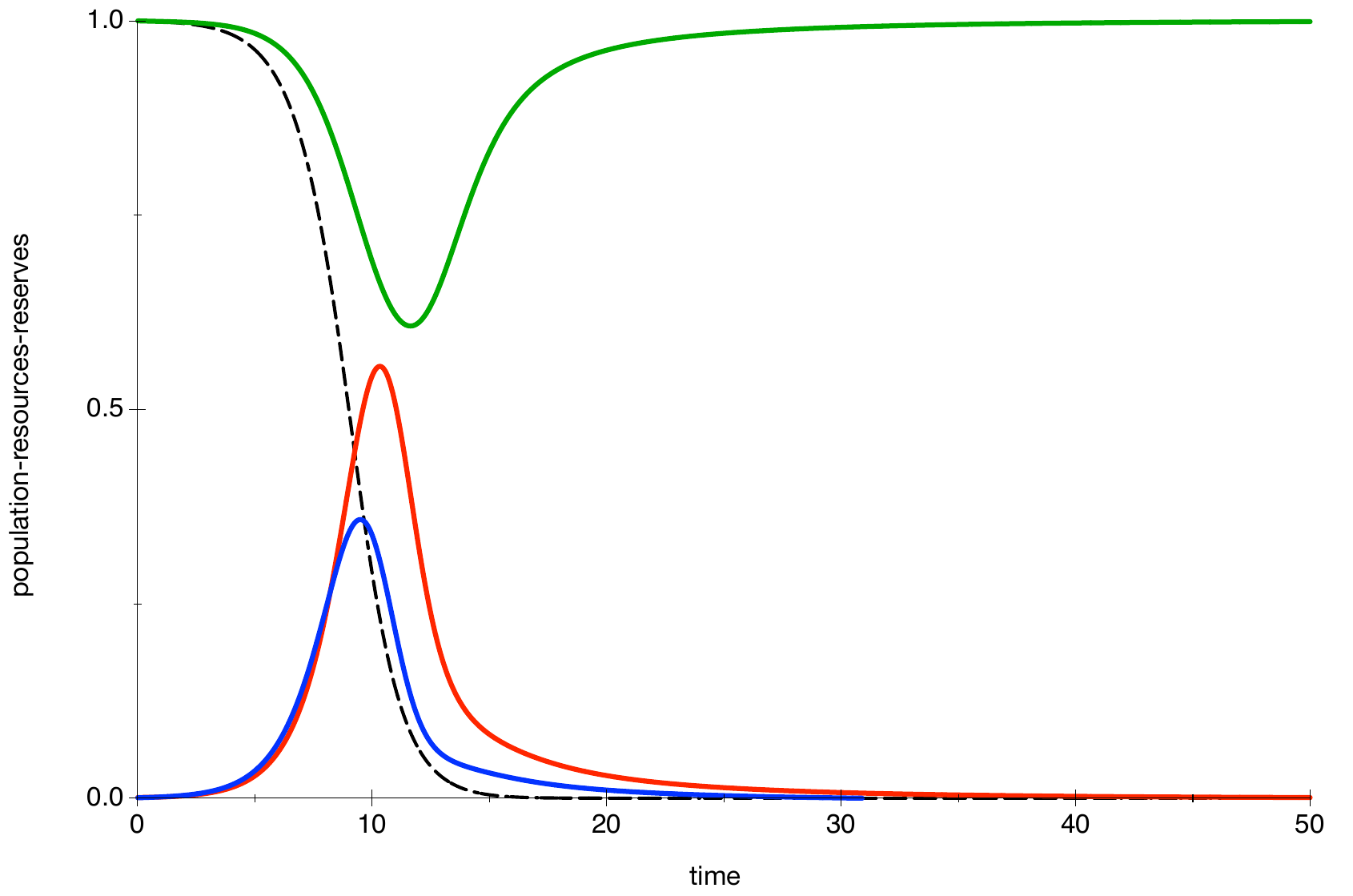}}}
\vskip-.25cm\qquad\centerline{\vbox{\hsize=14cm
\noindent Figure \three. A case of very slow collapse.
Simulation results for (\zdyo) with parameter values $\alpha=1, \beta=3, \lambda=2, \gamma=1.05, \delta=0, c=1, g=\tilde g=1, h=1$, starting from initial conditions $x_0=10^{-3}, y_0=1, r_0= \log(\beta/\alpha)/\lambda$ and $w_0=1$. The plots show 5000 iterations for a time step $\tau=10^{-2}$.}}
\smallskip

When the consumption parameter $\gamma$ is smaller than the extraction coefficient $h$, the collapse fixed point we studied above becomes unstable and the system reaches a steady state. Such a situation is presented in Fig.  \figdef\four.
\medskip
\centerline{\resizebox{10cm}{!}{\includegraphics{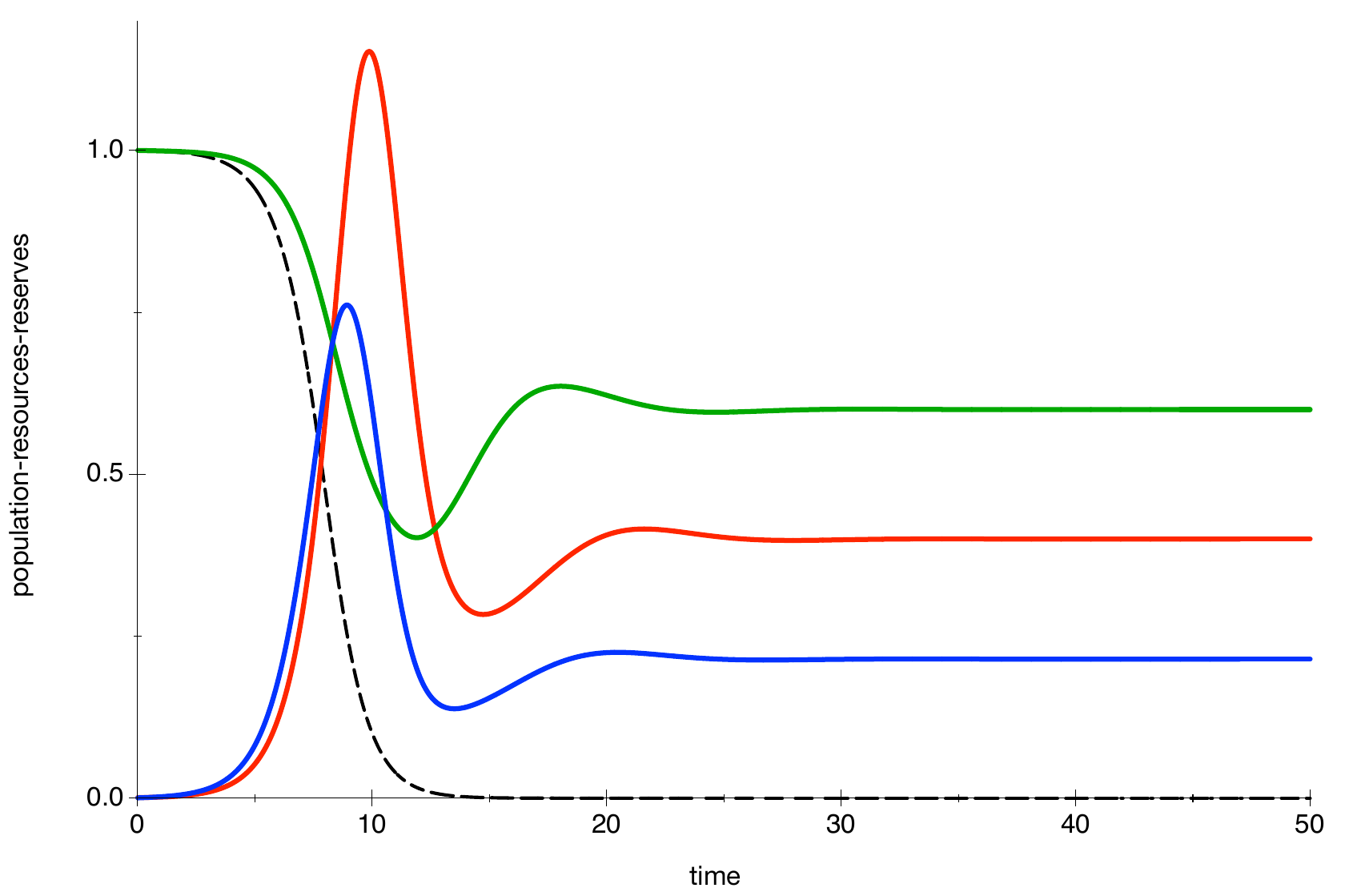}}}
\vskip-.25cm\qquad\centerline{\vbox{\hsize=14cm
\noindent Figure \four.  A case where the system reaches a steady state. Simulation results for (\zdyo) with parameter values $\alpha=1, \beta=3, \lambda=2, \gamma=0.6, \delta=0, c=1, g=\tilde g=1, h=1$, starting from initial conditions $x_0=10^{-3}, y_0=1, r_0= \log(\beta/\alpha)/\lambda$ and $w_0=1$. The plots show 5000 iterations for a time step $\tau=10^{-2}$.}}

\smallskip
We notice that the fixed point is reached after a small transient (which is not present if the value of $\gamma$ is very close to, but still smaller than, that of $h$). 
The existence of this oscillatory regime is particularly interesting and in order to be able to explore it fully we allow full freedom for $B(r)$ i.e. $\delta\ne0$. Fig. \figdef\five\ shows a situation where all variables oscillate strongly. 

\medskip
\centerline{\resizebox{10cm}{!}{\includegraphics{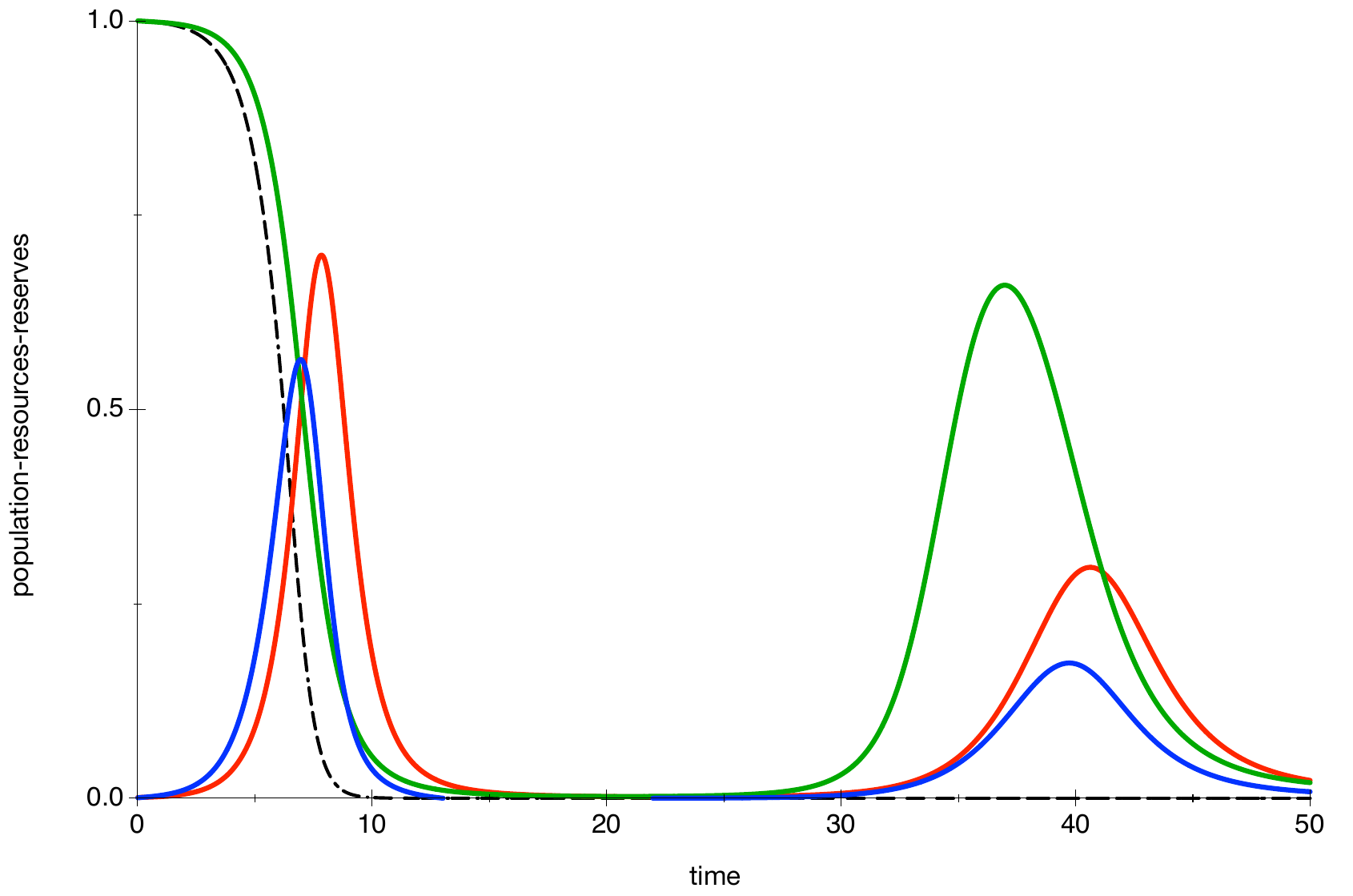}}}
\vskip-.25cm\qquad\centerline{\vbox{\hsize=14cm
\noindent Figure \five. A case of resurgence. Simulation results for (\zdyo) with parameter values $\alpha=1, \beta=3, \lambda=2, \gamma=2.1, \delta=2, \mu=1.5,  c=1, g=\tilde g=2, h=2.13$, starting from initial conditions $x_0=10^{-3}, y_0=1, r_0= \log(\beta/\alpha)/\lambda$ and $w_0=1$. The plots show 5000 iterations for a time step $\tau=10^{-2}$.}}

\smallskip
It is interesting to understand how this situation arises. First a rather large value of $g$ rapidly depletes the non-renewable resources while large values of both $h$ and $g$ fuel a population increase through the intermediary of $r$.  The renewable resources diminish substantially but the very low value of the subsistence minimum, $\gamma-\delta$, makes survival possible with scant resources, be it with a very small population. The renewable resources recover, followed by the population, but the tendency to a surplus consumption (large $\delta$) leads again to their dwindling, hence the repeated oscillations. And in fact, we are in a situation where the oscillations repeat ad infinitum, the system entering a limit cycle. A longer evolution clearly showing the convergence to the limit cycle is presented in Fig. \figdef\six.
\medskip
\centerline{\resizebox{10cm}{!}{\includegraphics{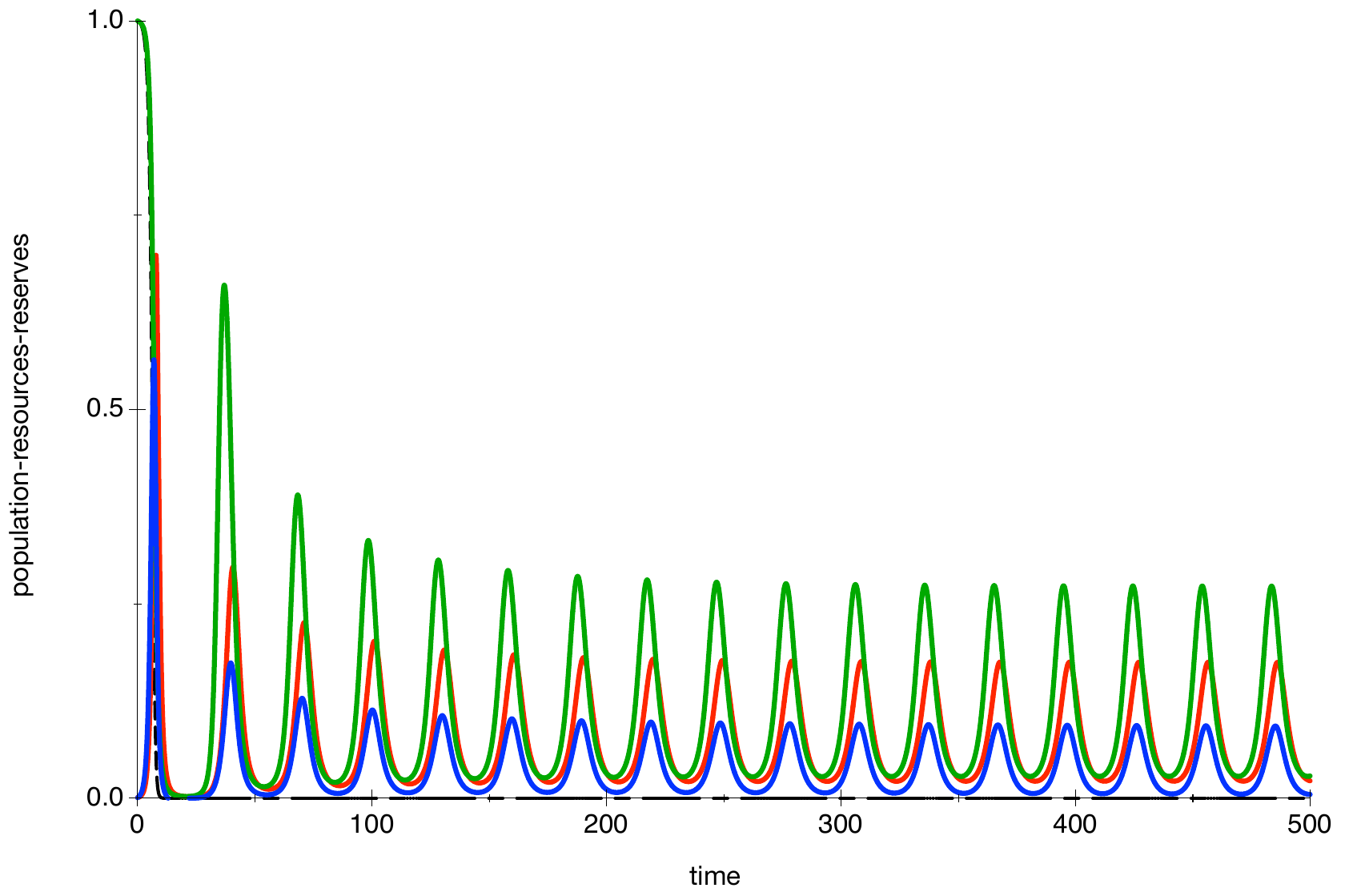}}}
\vskip-.25cm\qquad\centerline{\vbox{\hsize=14cm
\noindent Figure \six. The system enters a limit cycle. Simulation results for system (\zdyo) with the same parameter values as in Fig.  5 but for 50000 iterations.}}
\smallskip
In all the simulation results we have presented till now, the reserves, though dangerously ebbing in some cases, have always remained non-zero. This is not always the case. Going back to a standard case of collapse like the one presented in Fig. \one, and taking $g=2.$ instead of 1., leads to a complete depletion of reserves. The evolution based on (\zdekbis) with these parameters, leads in fact to negative values for $z$ and $r$ (which we put to 0, as previously explained). What is more interesting is a case where no collapse is predicted. In Fig. \figdef\seven\ we show such a case.
\medskip
\centerline{\resizebox{10cm}{!}{\includegraphics{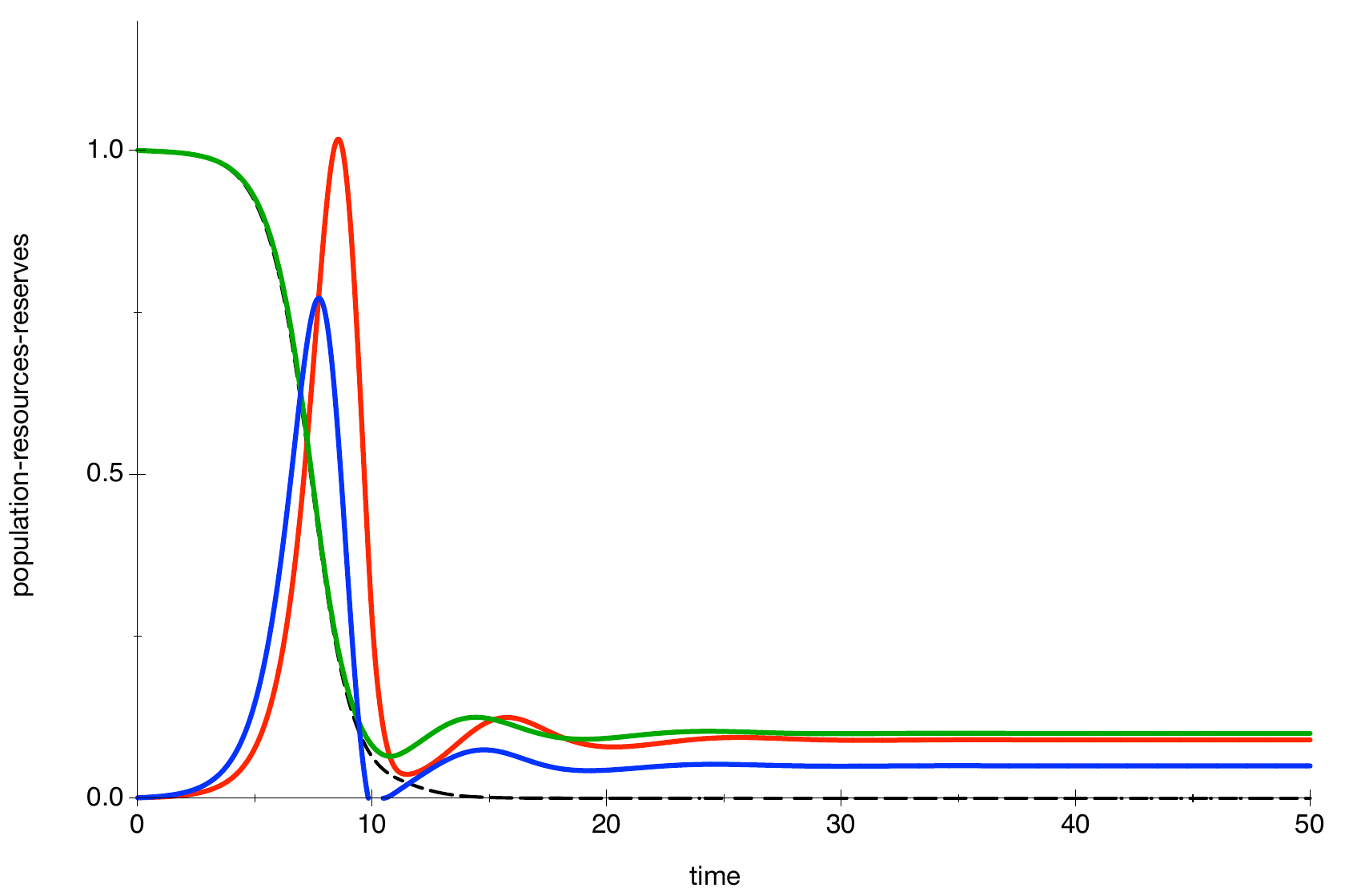}}}
\vskip-.25cm\qquad\centerline{\vbox{\hsize=14cm
\noindent Figure \seven. The reserves are exhausted at around $t=10$, but the system still manages to reach a steady state. Simulation results for (\zdyo) with parameter values $\alpha=1, \beta=3, \lambda=2, \gamma=1, \delta=0, c=1, g=\tilde g=1, h=1.9$, starting from initial conditions $x_0=10^{-3}, y_0=1, r_0= \log(\beta/\alpha)/\lambda$ and $w_0=1$. The plots show 5000 iterations for a time step $\tau=10^{-2}$.} }
\smallskip
During the evolution at one point the reserves are totally exhausted, i.e. we temporarily fix $z=0$ and $r=0$ and pursue the evolution. Clearly the value $r=0$ entails a decline of the population but the remaining population continues to extract from both renewable and non-renewable resources and at a certain point reserves again reach positive values. From then onwards the evolution proceeds normally leading to steady state with non-zero fixed point.

A still more interesting case is presented in Fig. \figdef\eight. 
\medskip
\centerline{\resizebox{10cm}{!}{\includegraphics{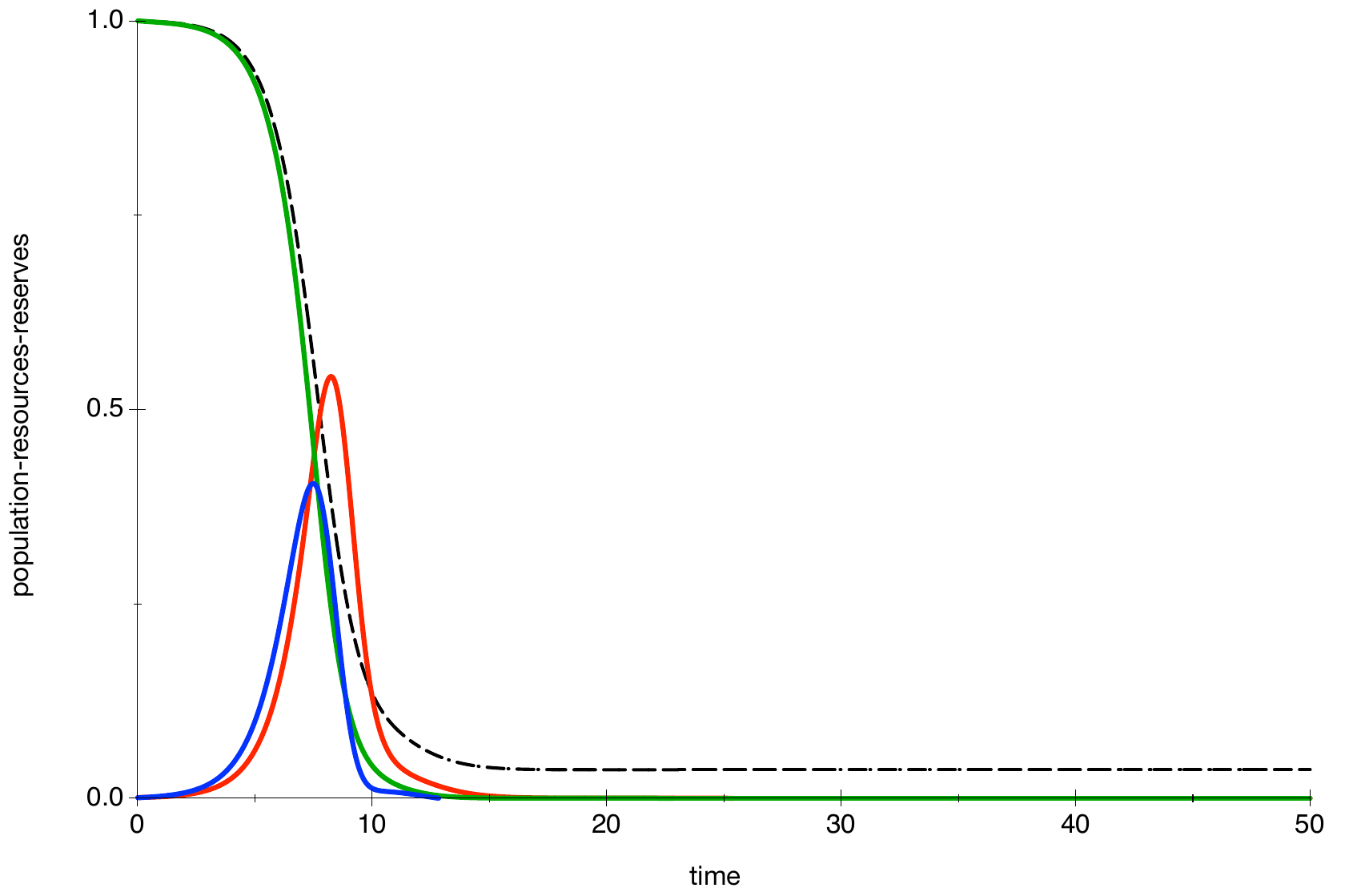}}}
\vskip-.25cm\qquad\centerline{\vbox{\hsize=14cm
\noindent Figure \eight. The population collapses and the renewable resources are not replenished. Simulation results for system (\zdyo) for $\alpha=1, \beta=3, \lambda=2, \gamma=2., \delta=1, \mu=2, c=1, g=\tilde g=1$ and $h=2.5$, starting from initial conditions $x_0=10^{-3}, y_0=1, r_0= \log(\beta/\alpha)/\lambda$ and $w_0=1$. The plots show 5000 iterations for a time step $\tau=10^{-2}$.} }
\smallskip
Although the steady state conditions are not violated, still the system collapses. But contrary to the ``standard'' collapse situation the renewable resources are not replenished and remain zero in the long run. Here the population goes to zero early enough, before exhausting the non-renewable resources. 

The fact that in this example the reserves are exhausted is not crucial, as we show in Fig. \figdef\nine\ where the parameters are chosen so that the reserves barely avoid crossing zero.
\medskip
\centerline{\resizebox{10cm}{!}{\includegraphics{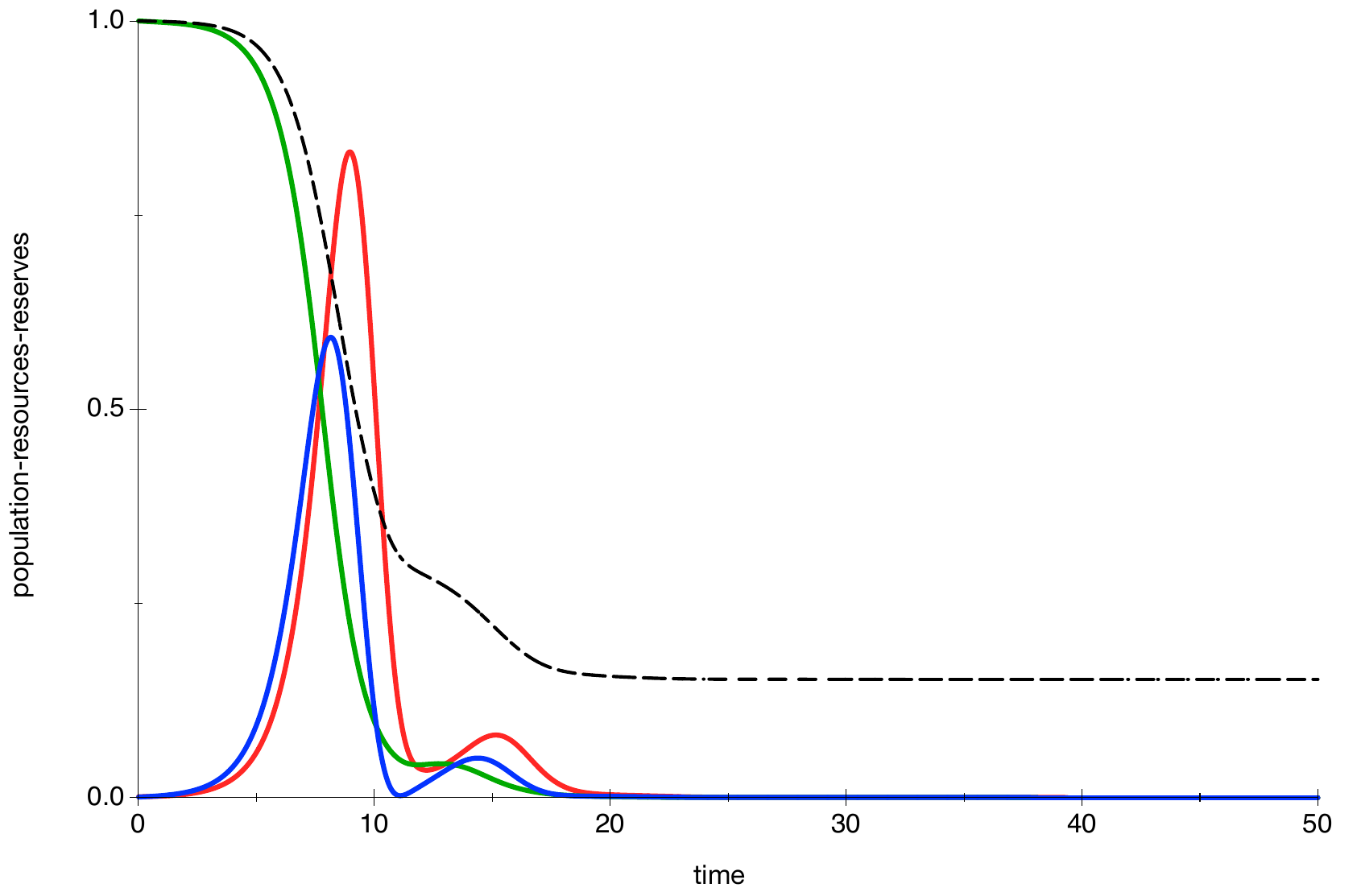}}}
\vskip-.25cm\qquad\centerline{\vbox{\hsize=14cm
\noindent Figure \nine. The population collapses before exhausting all non-renewable resources. Simulation results for system (\zdyo) for $\alpha=1, \beta=3, \lambda=2, \gamma=1., \delta=0, g=\tilde g=0.5$ and $h=2$, starting from initial conditions $x_0=10^{-3}, y_0=1, r_0= \log(\beta/\alpha)/\lambda$ and $w_0=1$. The plots show 5000 iterations for a time step $\tau=10^{-2}$.} }
\smallskip
Here, the reserves temporarily rebound, thanks to the use of non-renewable resources, but the level of the renewable ones is so low that they go to zero  and are not replenished in the long run. This results in a short-lived rebound and a total collapse occurs without total exhaustion of non-renewable resources. 
\bigskip
5. {\scap Societal effects}
\medskip
The term `societal’, as used in this paper's title and the present section, calls for elucidation. Here, it denotes the potential influence of society on the evolution of population and resource reserves. This influence can manifest itself in two distinct ways: passive or active.

Passive influence encompasses all effects that humanity exerts on its natural environment, by its mere presence in that environment. The repurposing of land, the pollution of the atmosphere, oceans, and inland waters, the degradation of flora and fauna are characteristic examples of such passive societal effects. This in the sense that they are, for the most part, unintended. In our model, we shall account for such effects, globally, by introducing a nature capacity $C(x)$ that is a decreasing function of the population.
Another form of passive societal influence arises from inefficiency and waste. In Sec. 2, where we presented our model, we introduced the coefficient $\tilde g$ to represent the extraction of non-renewable resources, with the understanding that it can be larger than the coefficient $g$ that is involved in the creation of reserves. This excess, reflecting waste and inefficiency, is inherent in the functioning of human society and is, for the most part, non-deliberate. A similar inefficiency likewise affects the extraction of renewable resources. When formulating our model, we chose not to incorporate this effect and applied an appropriate scaling to the dynamical variables instead. This allowed us to conduct the full stability analysis of the model without additional complexity. 

Active influence is almost self-explanatory. It encompasses the deliberate actions undertaken by human societies to modify the evolution of the population and of the resources and reserves. There is however a well-known behavioral tendency to overlook early warning signals, to maintain existing practices, and to respond only (if at all) when the prospect of collapse becomes imminent. Consequently, in the study of active societal effects, it is of limited relevance to examine early interventions, since timely measures would evidently prevent collapse. Of greater interest is whether, when the system approaches a critical state, effective intervention can still mitigate or even avert collapse. How, then, can such measures be implemented? One possible mechanism is through adjustments in the production rates of reserves derived from renewable and non-renewable resources. Another is through modifications in consumption dynamics, reflecting changes in societal demand. This final point requires some clarification. In several of the simulations presented in Section 4, we adopted a consumption term $B$ that is independent of the per capita reserves $r$ (by setting $\delta=0$). In Section 2, we referred to this case as `mere subsistence' and `baseline necessity'. However, this should not be understood as representing the physiological subsistence level of human beings, which is essentially fixed by biological constraints and therefore incompressible. The baseline in our formulation is instead the societal threshold required for the proper functioning of social structures. In this sense, considering a reduced baseline remains conceivable: it would effectively correspond to lowering expenditures on collective services such as health, education, or security. While such a reduction might entail the loss or hardship of one or more generations, our focus here is on mathematical modelling rather than policy prescription. From this perspective, exploring such a measure is justified if its inclusion would allow the system to avoid global collapse.

We start our presentation of results by revisiting the case that gave rise to Fig. \eight. As a first measure to avert collapse we diminish the renewable resources extraction coefficient $h$ from 2.5 to 1.4 at time $t=13$, i.e. after 1300 iterations performed with the initial value of $h$. As shown in Fig. \figdef\ten\ below, this does not suffice in order to thwart a breakdown. 
\medskip
\centerline{\resizebox{10cm}{!}{\includegraphics{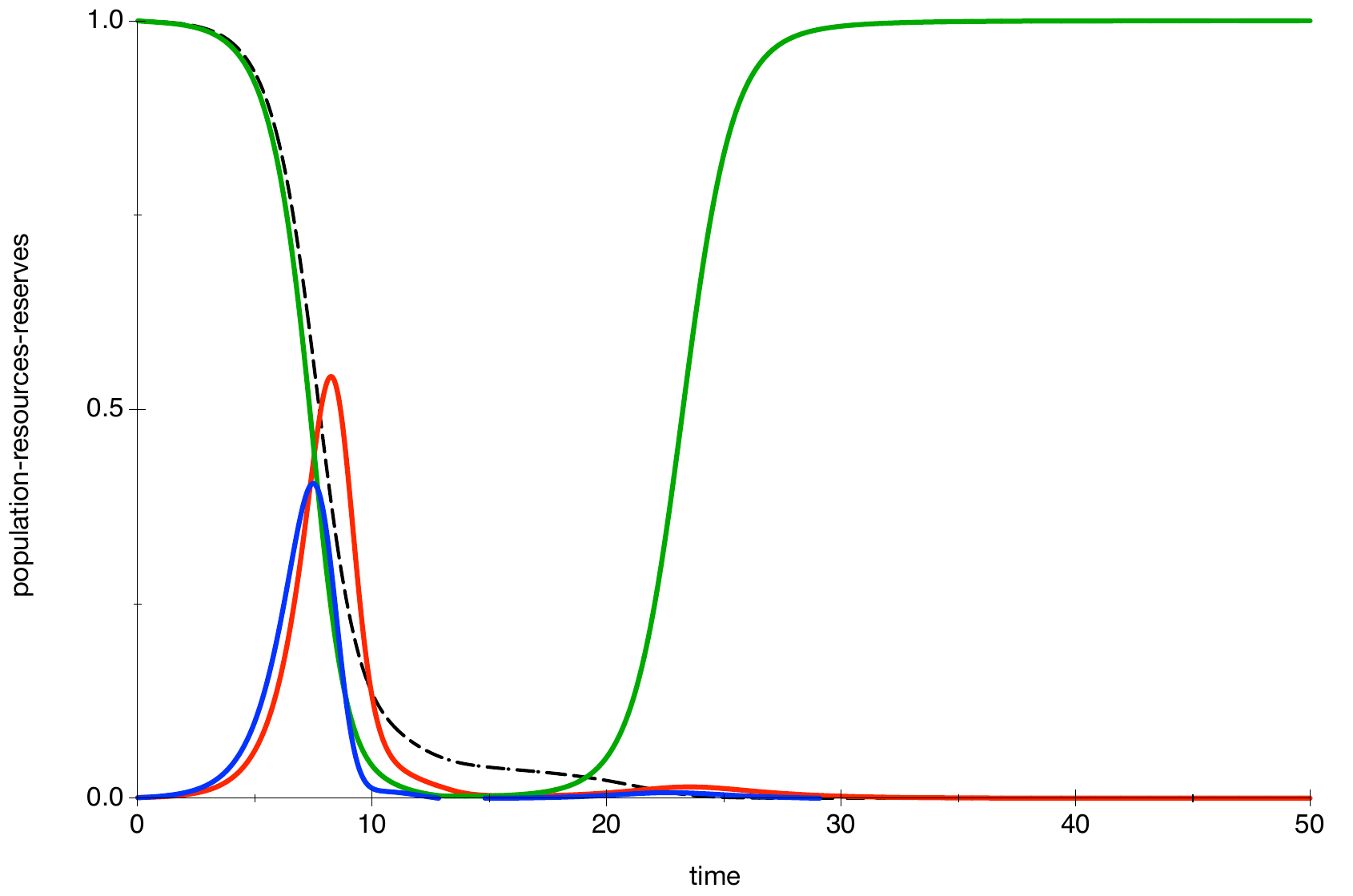}}}
\vskip-.25cm\qquad\centerline{\vbox{\hsize=14cm
\noindent Figure \ten. A change of resource extraction at $t=13$ allows the renewable resources to be replenished but the population still collapses. Simulation results for system (\zdyo) for $\alpha=1, \beta=3, \lambda=2, \gamma=2, \delta=1, \mu=2, g=\tilde g=1$ and $h=2.5$ initially, then $h=1.4$ as of $t=13$. The plots show 5000 iterations for a time step $\tau=10^{-2}$, starting from initial conditions $x_0=10^{-3}, y_0=1, r_0= \log(\beta/\alpha)/\lambda$ and $w_0=1$.} }
\smallskip
Renewable resources are eventually replenished, allowing the population to persist much longer, but, as the consumption rate remains high, extinction turns out to be inevitable in the long run. So as a next attempt at forestalling collapse we introduce, along with the above reduction of renewable resources extraction, a massive reduction of consumption by putting $\delta=0$ and reducing $\gamma$ to a subsistence level of 1, again at $t=13$. This time these measures are suffcient, as can be assessed from Fig. \figdef\eleven. 
\medskip
\centerline{\resizebox{10cm}{!}{\includegraphics{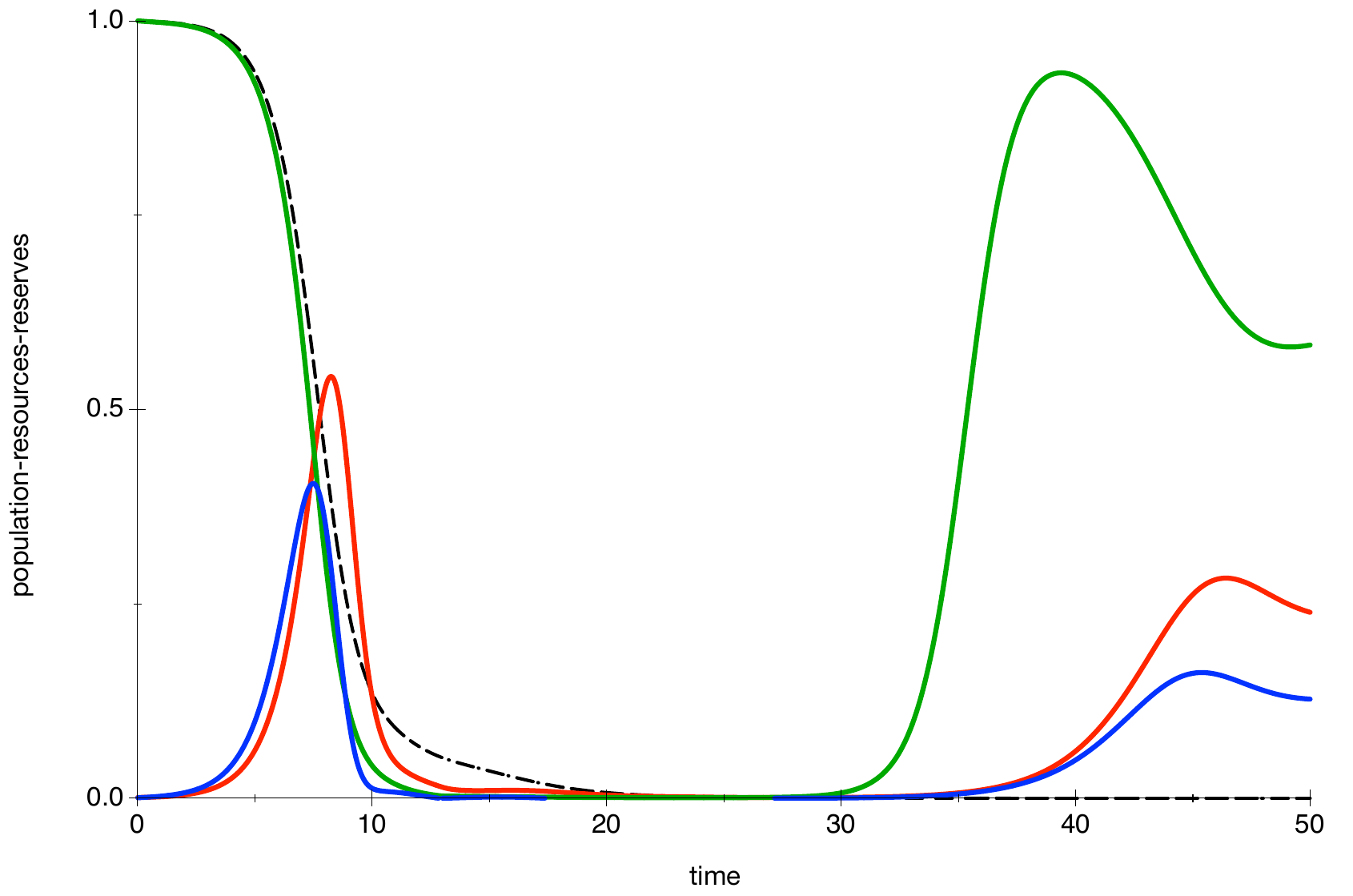}}}
\vskip-.25cm
\qquad
\centerline{\vbox{\hsize=14cm
\noindent Figure \eleven. Diminishing resource extraction as well as consumption allows one to avoid total collapse. Simulation results for system (\zdyo) for $\alpha=1, \beta=3, \lambda=2, \gamma=2, \delta=1, \mu=2, g=\tilde g=1$ and $h=2.5$ initially, then as of $t=13$,  $\gamma=1, \delta=0$, and $h=1.4$. The plots show 5000 iterations for a time step $\tau=10^{-2}$, starting from initial conditions $x_0=10^{-3}, y_0=1, r_0= \log(\beta/\alpha)/\lambda$ and $w_0=1$.} }
\smallskip
We remark that once the measures are implemented the system reaches a steady state. Note however that not only the precise parameter values involved in this type of intervention, but also the precise point of intervention is crucial for the eventual outcome and that the interplay between parameters and intervention time seems highly nontrivial. Had we postponed the intervention by a small time interval or assumed a slightly larger value of $h$, the recovery would have been impossible.

We turn now to the case shown in Fig. \nine\ where the reserves are exhausted and, contrary to the case presented in Fig. \seven, they are never sufficiently replenished with the population eventually collapsing. In Fig. \figdef\twelve\ we present the evolution of the system for the same parameter values as in Fig.  \nine\   but where at time $t=11$, after 1100 iterations with a value of $h=2.$, we adjust the extraction coefficient for the renewable resources to the lower value $h=1.2$. 
\medskip
\centerline{\resizebox{10cm}{!}{\includegraphics{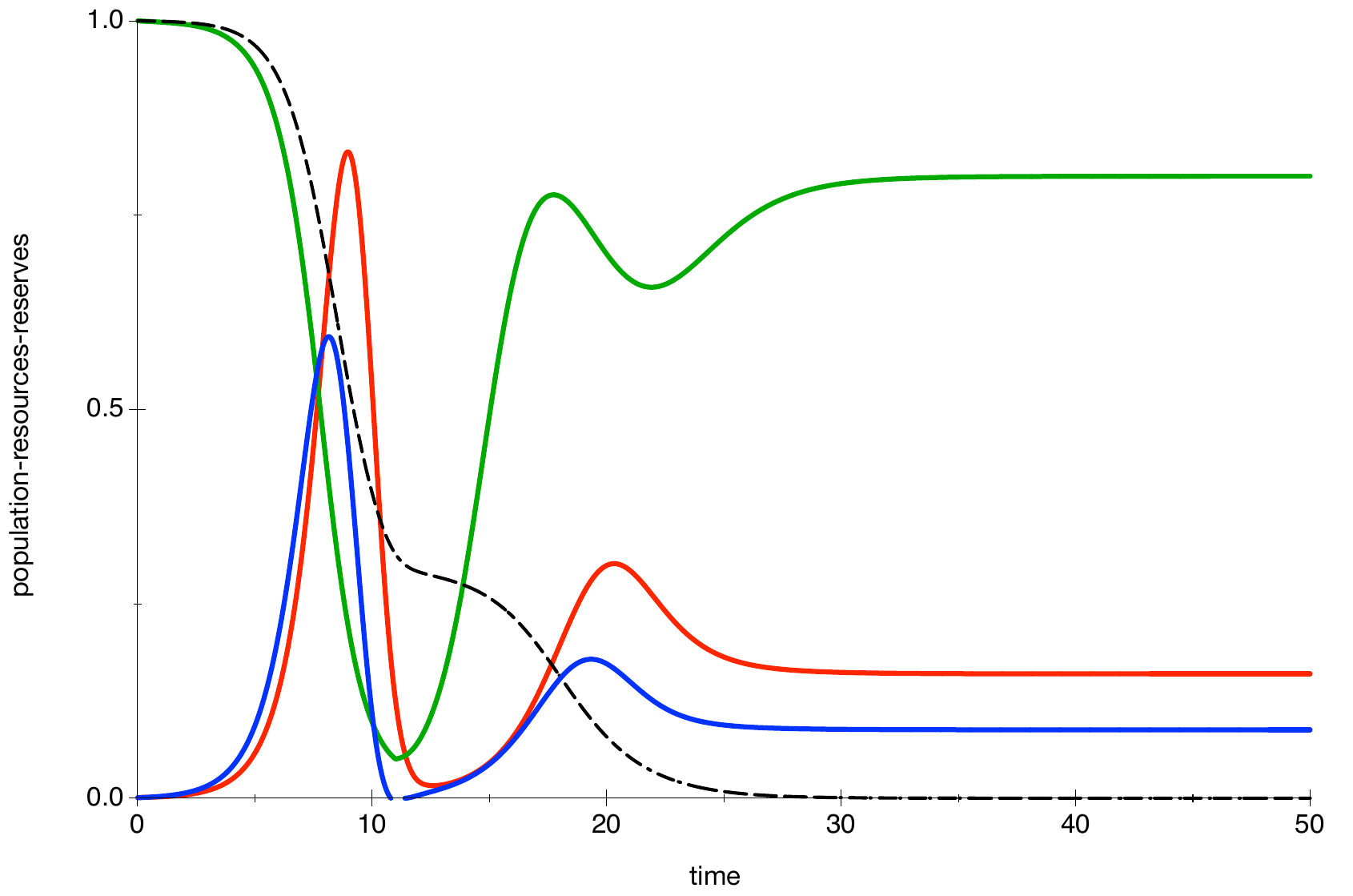}}}
\vskip-.25cm
\qquad
\centerline{\vbox{\hsize=14cm
\noindent Figure \twelve. Diminishing renewable resource extraction allows one to avoid collapse despite reserves approaching zero. Simulation results for system (\zdyo) for $\alpha=1, \beta=3, \lambda=2, \gamma=1., \delta=0, g=0.55$ and $h=2$ initially, then $h=1.2$ at $t=11$. The plots show 5000 iterations for a time step $\tau=10^{-2}$, starting from initial conditions $x_0=10^{-3}, y_0=1, r_0= \log(\beta/\alpha)/\lambda$ and $w_0=1$.} }
\smallskip
Since the consumption coefficient was already fixed at a subsistence-level value, with $\delta=0$, 
the effect of lowering the value of $h$ at $t=11$ is that the population does not present the temporary rebound observed in Figure \nine\ around $t=11$ but diminishes even further, which lowers the burden on the renewable resources and allows a recovery, leading eventually to a steady state.

We turn now to effects that we call `passive'. As explained in section 4,  an inefficient extraction of non-renewable resources ($\tilde g> g$) does not  have a decisive  effect. 
The main `passive' effect we are going to study here is that of the degradation of nature, resulting in a reduction of the capacity $C$. We shall assume that the capacity varies with population as
$$C(t)={1\over 1+f\, x(t-T)},\eqdef\dena$$
with $f$ a non-negative constant (where we have normalised the capacity to 1, when $f=0$, as explained in section 2) and a delay time $T\geq 0$. Three different scenarios are explored. In the first, we assume that the response of nature to the pressure exerted by the population $x(t)$ is instantaneous, i.e. that $T=0$ and that $C(t)$ follows indeed the population at every instant $t$. In the second scenario we assume that this response involves some delay, i.e. $T>0$. In the third scenario we make the extreme assumption that nature never recovers: the capacity diminishes up to the point where the population reaches a (first) maximum and then forever remains at that same minimum. These scenarios are admittedly extreme---instantaneous response is impossible, as is an absence of recovery---but we believe they effectively bracket more realistic cases.

The case of zero-recovery is the worst case scenario. For instance, if one considers a situation like the one of Fig. \four\ which normally leads to a steady state, a capacity diminishing and never rebounding may lead to complete extinction for values of $f$ larger than 1.5. Similarly, a limit cycle disappears for all values of $f>0.1$. The remaining two scenarios do preserve the dynamical behaviour of the system with $C=1$, shown in Fig. \four, perhaps going through an intermediate phase of total depletion of reserves (for large values of $f$). 

The really interesting case is the one associated with the parameters that give rise to Fig. \seven. We compare the behaviour of the three scenarios for $f=0.5$. The case of no recovery is shown in Fig. \figdef\fourteen.  
\medskip
\centerline{\resizebox{10cm}{!}{\includegraphics{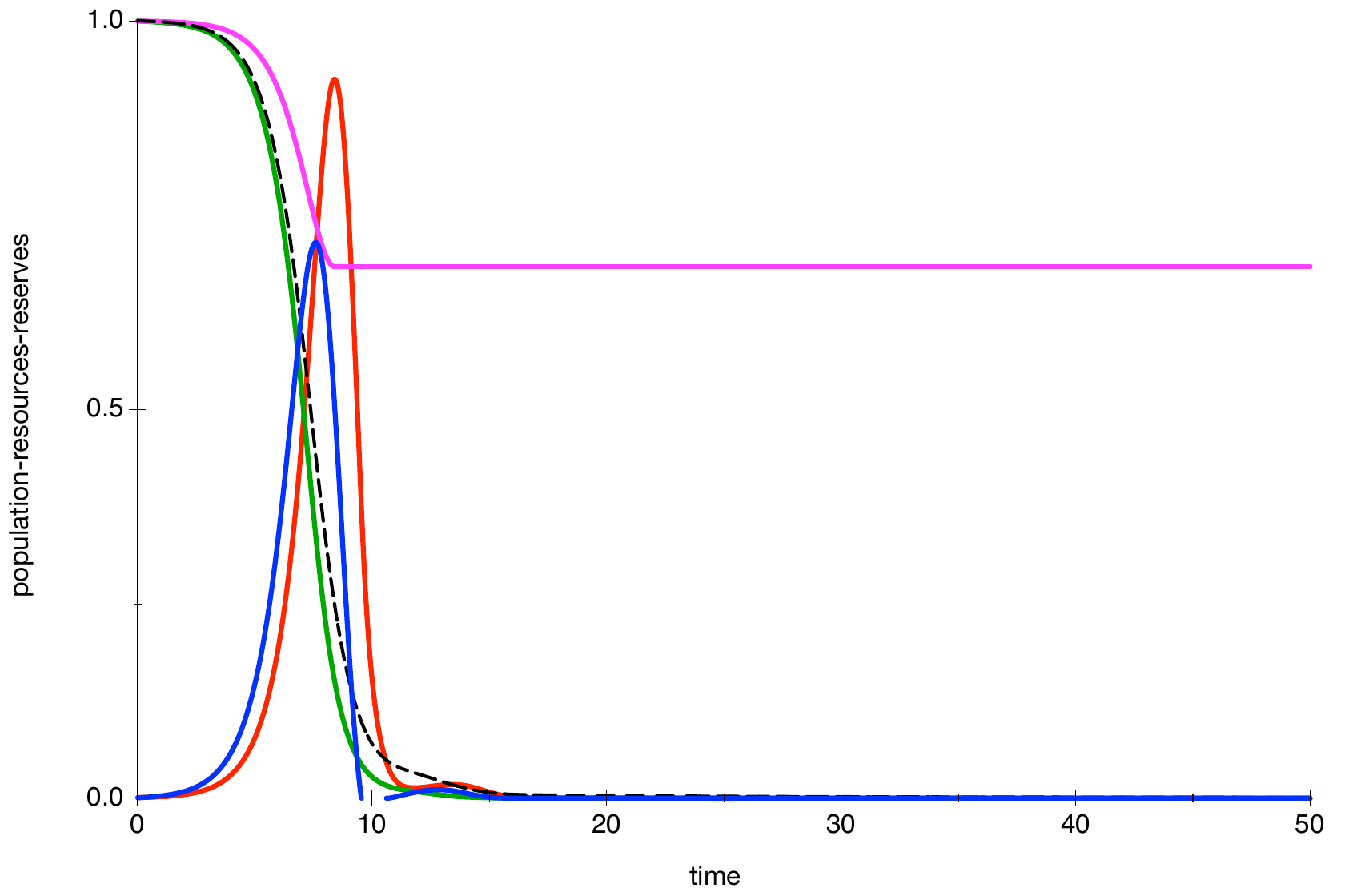}}}
\vskip-.25cm
\qquad
\centerline{\vbox{\hsize=13cm
\noindent Figure \fourteen. Permanent damage to nature invariably leads to collapse. Simulation results for (\zdyo) with parameter values $\alpha=1, \beta=3, \lambda=2, \gamma=1, \delta=0, c=1, g=\tilde g=1, h=1.9$ and $f=0.5$ in $C(x)$ of the form (\dena), starting from initial conditions $x_0=10^{-3}, y_0=1, r_0= \log(\beta/\alpha)/\lambda$ and $w_0=1$, but with $C(x)$ remaining constant after it reaches a minimum. The plots show 5000 iterations for a time step $\tau=10^{-2}$. The pink line depicts the state of the capacity $C(x(t))$. }}
\smallskip
We see that the decreased value of the capacity suffices for recovery to be impossible. 

The behaviour of the system in the two remaining scenarios is similar. In the case of instant recovery the decreasing capacity has as effect an earlier exhaustion of reserves and a diminution of the population. However this is not sufficient for a collapse to occur and eventually the population recovers, reaching a steady state. The other case is that of the delayed reponse. This scenario is different from the previous two, since it introduces one more parameter, that of the delay $T$. Given the characteristic times observed in the growth and decrease of the population, a reasonable value of $T$ would be of the order of a few hundred iterations. We opt for a value of $T=2.5$ (and we verified that the results do not change appreciably when we increase this value up to 5). Fig. \figdef\fifteen\ shows the results of a simulation with a delay of 250 iteration steps.
\medskip
\centerline{\resizebox{10cm}{!}{\includegraphics{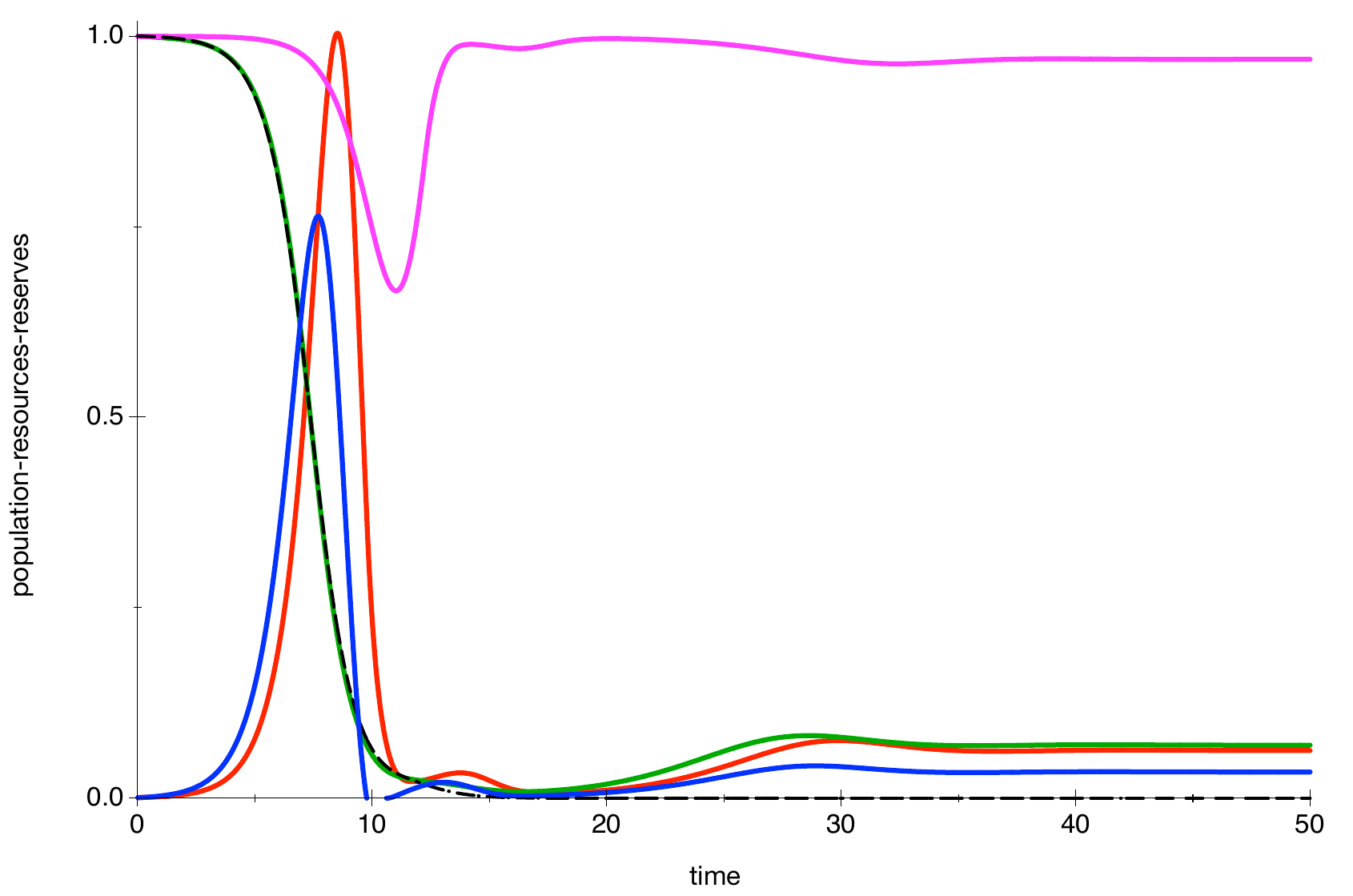}}}
\vskip-.25cm
\qquad
\centerline{\vbox{\hsize=13cm
\noindent Figure \fifteen. Delayed damage (and revovery) of nature allows to reach steady state despite temporary reserves' exhaustion. 
Simulation results for (\zdyo) with parameter values $\alpha=1, \beta=3, \lambda=2, \gamma=1, \delta=0, c=1, g=\tilde g=1, h=1.9$, and $f=0.5$ and $T=2.5$ in $C(x)$ of the form (\dena), starting from initial conditions $x_0=10^{-3}, y_0=1, r_0= \log(\beta/\alpha)/\lambda$ and $w_0=1$, but with $C(x)$ remaining constant after it reaches a minimum. The plots show 5000 iterations for a time step $\tau=10^{-2}$. The pink line depicts the state of the capacity $C(x(t))$. }}\smallskip
We observe a recovery of the population, very similar to the case of instant response, the small differences between the two scenarios being confined to the intermediate region between massive decrease and resurgence.

A question that naturally arises at this point is what happens when the response of capacity to population is delayed but without possibility of recovery. We have verified, in the same conditions as in the examples just above, that the permanent reduction of capacity, be it delayed, suffices to prevent any resurgence of the population. However, this serves merely as an illustration of a possible scenario; the detailed behavior of the system will depend on the precise values of the parameters involved. 
\bigskip
6. {\scap Conclusion}
\medskip
Predicting the future has long preoccupied humanity, across civilisations. Since antiquity, various methods of divination have been devised, achieving success only equal to pure chance amid unknown probabilities. Spurious correlations were elevated to predictors of the future, fostering a class of specialists, from ancient shamans to modern charlatans, who thrive on the pervasive illusion that the future can be accurately foreseen (only by them, of course). Compiling historical data and subjecting them to statistical analysis enabled more informed forecasts, though the validity of such forecasts is fundamentally no greater than that of probability estimates. The advent of mathematical modelling, however, marked a turning point. The most prominent example is weather forecasting, where models of atmospheric dynamics, coupled with extensive and timely data collection, yield reliable, though short-term, meteorological predictions. But what about long-term forecasts? We are convinced that mathematical modelling is the way forward here too, provided the questions posed are suited to such forecasts, eschewing the temptation for detailed, precise, quantitative predictions. This means we expect modelling approaches to reveal tendencies in future evolution, possible outcomes, and---at best---the key parameters driving them. This is the approach adopted in the present work.

In this paper, we address the potential disappearance of our present civilisation, amid extinction prognostications by various futurologists. This is hardly a novel question: even setting aside doom prophecies that have shadowed humanity through the ages, the survival of human society has been the subject of mathematical modelling studies for over 50 years. Forrester and the MIT team led by Meadows were the first to propose a model using the former's `system dynamics' approach [\preforrester]. Their studies concluded that most evolutionary scenarios lead to societal collapse from overpopulation and resource scarcity. However, the complexity of models like Forrester's, with hundreds of dependent variables and thousands of parameters, in our view undermines the robustness of such conclusions. To circumvent this issue, we adopted a simple model, originally introduced under the HANDY moniker in a different context, that of the survival of a multi-layered society [\originalhandy]. We simplified this model even further by assuming a single, homogeneous population and investigated the conditions under which such a system leads to either population and resource collapse or, conversely, to a steady-state equilibrium.

In [\notrehandy], we presented an initial study of the single-population HANDY model, emphasising its dynamical systems aspects. We investigated appropriate interaction terms drawing from Holling’s predator-prey theories, introduced a suitable discrete analogue of the differential system, and examined the ‘ultra-discrete’ limit as the discretisation parameter grows large. However, that preliminary work inadequately addressed non-renewable resources and relied on a highly specific positivity-preserving discretisation (which was needed to be able to address the ultradiscrete limit of the system). In the present model we incorporate non-renewable resources via an ad-hoc dynamical variable, while positivity of reserves is ensured through an appropriate interpretation of the evolution rules. Moreover, we accounted for the impact of population pressure on the environment---referred to as societal effects---and in particular the reduction in Nature's regenerative capacity caused by human presence [\refdef\giec], which leads to a lower availability of renewable resources. Finally, special interaction terms---different from those of {[\cryptic] but also based on Holling's theories---were introduced for the terms modelling the extraction of renewable and non-renewable resources.

The dynamics of our variant of the HANDY system were studied through extensive numerical simulations, leading to two distinct regimes. The first is that of a collapse where both population and reserves tend to zero. Depending on the parameters, this collapse can either correspond to a rapid extinction or to a protracted decline. The second regime is a steady state, which, in some cases, can be reached after an oscillatory intermediate state. The presence of such oscillations can, under certain special conditions, give rise to a permanently oscillating, limit-cycle-type regime. The two kinds of resources, renewable and non-renewable, do not play the same role in the dynamics, with the former type being particularly crucial. In our simulations we observed that, in some cases, the reserves can reach zero. Continuing the evolution under a non-negativity constraint, i.e., fixing the value at zero, in some case allows the system  to rebound and even to reach a steady state. We also found that changing policies, even late in the evolution, may in some cases alter the outcome and prevent collapse. And what is particularly interesting, as shown in Fig. \figdef\sixteen, is that it is even possible, once the situation has stabilised, to go back to the previous level of production and consumption without inducing a collapse: the initial `culling' of the population during the quasi-collapse, combined with the  lower population level that results from the absence of non-renewable resources, suffices for this. 
\medskip
\centerline{\resizebox{10cm}{!}{\includegraphics{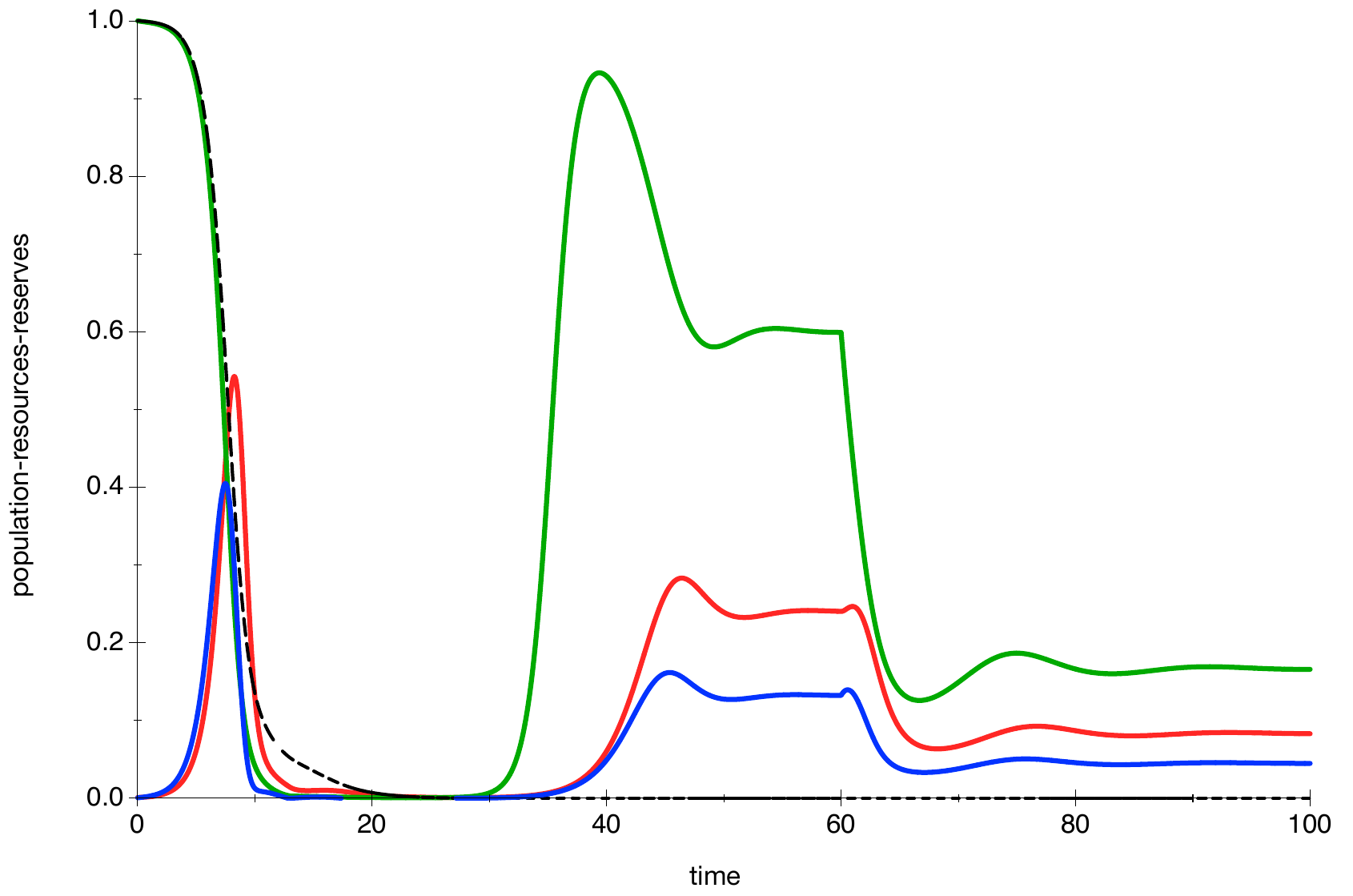}}}
\vskip-.25cm
\qquad
\centerline{\vbox{\hsize=13cm
\noindent Figure \sixteen. Diminishing resource extraction and consumption allows to avoid total collapse, an effect which is robust despite a later backsliding to old policies. Simulation results for system (\zdyo) for $C\equiv1, \alpha=1, \beta=3, \lambda=2, \gamma=2, \delta=1, \mu=2, g=1$ and $h=2.5$ initially, then as of $t=13$,  $\gamma=1, \delta=0$, and $h=1.4$ and return to the initial parameter values as of $t=60$. The plots show 10000 iterations for a time step $\tau=10^{-2}$, starting from initial conditions $x_0=10^{-3}, y_0=1, r_0= \log(\beta/\alpha)/\lambda$ and $w_0=1$.} }
\smallskip
Finally, investigating societal effects, we remarked that when the damage to nature’s regenerative capacity is permanent, the collapse of population and reserves is often unavoidable.

What is the main take-away of this study? First, we can remark that a very simple, fourth-order differential system, such as the variant of HANDY that we analysed here, can display remarkably rich dynamical behaviour. This is proof, if ever there were a need for it, of the power of mathematical modelling when applied to real-life situations. On the other hand, our work highlights the crucial role of the various parameters of the system and their influence on the ultimate outcome. This should serve as a permanent warning against the use of ultra-complicated models involving hundreds of badly (or not at all) controlled parameters.
What, then, is the main conclusion concerning the survival or collapse of human society stemming from our study? (After all, this is the question we set out to address). While making predictions, that aim to be realistic, regarding the current state of the world, lies beyond the scope of this work, we do find that societal collapse is indeed a possibility. Moreover, if a resurgence occurs, the resulting society may be fundamentally different, due to the potential exhaustion of non-renewable resources---and very likely one in which computer simulations, such as those of the present study, would belong to the realm of the past, if not to mythology.

\bigskip
{\scap Potential sources of conflict}
\medskip
The authors declare that there are no potential sources of conflict involving the research results presented in this manuscript.

\bigskip
{\scap References}
\medskip
\item{[\raupsep]} D.M. Raup and J.J.J. Sepkoski, Mass extinctions in the marine fossil record, Science {\bf 215}  (1982), 1501--1503.
\item{[\bigfive]} C.R. Marshall, Forty years later: The status of the “Big Five” mass extinctions, Cambridge Prisms: Extinction. (2023) 1:e5.
\item{[\bolide]}  A.A. Chiarenza, A. Farnsworth A., P.D Mannion., D.J. Lunt,  P.J. Valdes, J.V. Morgan  and P.A. Allison, Asteroid impact, not volcanism, caused the end-Cretaceous dinosaur extinction, Proc. Natl. Acad. Sci. U.S.A. {\bf 117} (2020) 17084--17093.
\item{[\malthus]} T. Malthus, {\sl An essay on the principle of population}, J. Johnson, London, 1798.
\item{[\lancet]} S.E. Vollset et al., { Fertility, mortality, migration, and population scenarios for 195 countries and territories from 2017 to 2100: a forecasting analysis for the Global Burden of Disease Study}, Lancet, {\bf 396} (2020) 1285-1306.
\item{[\forrester]} J. Forrester, {\sl World Dynamics}, Wright-Allen Press, Cambridge,  Massachusetts, 1971.
\item{[\preforrester]} J. Forrester, {\sl Urban Dynamics}, MIT Press, Cambridge, Massachusetts, 1969.
\item{[\predicament]} The Club of Rome, {\sl The Predicament of Mankind: Quest for Structured Responses to Growing World-wide Complexities and Uncertainties}, A proposal, 1970.
\item{[\limits]} D. Meadows, J. Randers, D. Meadows and W. Behrens, {\sl  The Limits to growth: A report for the Club of Rome's Project on the Predicament of Mankind}, Universe Books, New York, 1972.
\item{[\supporters]} G. Turner, { A comparison of The Limits to Growth with 30 years of reality}, Global Environmental Change {\bf 18} (2008) 397-411.
\item{[\jackson]} T. Jackson and R. Webster, {\sl Limits Revisited--A review of the limits to growth debate. A report to the All-Party Parliamentary Group on Limits to Growth}, 2016, DOI:10.13140/RG.2.2.21095.91045.
\item{[\brandertaylor]} J. Brander and M.S. Taylor, { The simple economics of Easter Island:  A Ricardo Malthus model of renewable resource use}, The American Economic Review {\bf 88} (1998) 119-138.
\item{[\diamond]} J.M. Diamond, {\sl Collapse: How Societies Choose to Fail or Succeed} Penguin, 2006.
\item{[\rapanuipop]} D.S. Davis et al., Island-wide characterization of agricultural production challenges the demographic collapse hypothesis for Rapa Nui (Easter Island), Science Advances {\bf 10} (2024) eado1459.
\item{[\rapanuigen]} V. Moreno-Mayar J., et al., Ancient Rapanui genomes reveal resilience and pre-European contact with the Americas, Nature {\bf 633} (2024) 389--397.
\item{[\cryptic]} R. Willox, A. Ramani and B. Grammaticos, A discrete-time model for cryptic oscillations in predator-prey systems,  Physica D {\bf 22} (2009) 2238-2245.
\item{[\originalhandy]} S. Motesharrei, J. Rivas and E. Kalnay, {\sl Modeling inequality and use of resources in the collapse or sustainability of societies}, Ecol. Econ. {\bf 101} (2014) 90-102.
\item{[\notrehandy]} B. Grammaticos, R. Willox and J. Satsuma, Revisiting the Human and Nature Dynamics model, Reg. Chao. Dyn. {\bf 25} (2020) 178.
\item{[\akhavanyorke]} N. Akhavan and J. Yorke, { Population Collapse in Elite-Dominated Societies: A Differential Equations Model without Differential Equations}, SIAM J. App. Dyn. Sys. {\bf 19}, (2020), 1736-1757.
\item{[\shilorkadhim]} M. Shillor, T. A. Kadhim, { Analysis and Simulations of the HANDY Model
with Social Mobility, Renewables and Nonrenewables}, Electronic Journal of Differential Equations, Vol. 2023 (2023), No. 59, pp. 1-22.
\item{[\tonnelier]} A. Tonnelier, {\sl Sustainability or Societal Collapse: Dynamics and Bifurcations of the HANDY Model}, SIAM J. App. Dyn. Sys. {\bf 22}, (2023), 1977.
\item{[\patry]} L. Patry, P. Morel, E. Tomasi-Gustafsson, E. Kalnay, J. Rivas and S. Mote {\sl Modeling the effects of natural disasters, wars, and migrations on sustainability or collapse of pre-industrial societies: Random perturbations of the Human and Nature Dynamics (HANDY)}, preprint 2024, https://doi.org/10.48550/arXiv.2407.14860
\item{[\holling]} C. Holling, { Some Characteristics of Simple Types of Predation and Parasitism}, Canadian Entomologist {\bf 91} (1959) 385-398.
\item{[\gantmacher]} F.R. Gantmacher, {\sl Matrix Theory}, McGraw-Hill, N.Y., 1959, Volume 2, p. 185.
\item{[\covid]} G. Nakamura, B. Grammaticos, M. Badoual, { Confinement strategies in a simple SIR model}, Reg. Chao. Dyn. {\bf 25} (2020) 509-521.
\item{[\bardi]} U. Bardi, {\sl The Seneca Effect: Why Growth is Slow but Collapse is Rapid}, Springer N.Y. 2017.
\item{[\giec]}  M. Jones et al., {\sl IPBES Business and Biodiversity Assessment: Summary for Policymakers}, (2026), DOI 10.5281/zenodo.15369060.

\end